\documentclass[twocolumn,prd,aps,superscriptaddress,eqsecnum]{revtex4-1}
\usepackage[T1]{fontenc}
\usepackage[latin9]{inputenc}

\usepackage{amsmath,amssymb,bm,bbm}
\usepackage{dsfont}
\usepackage{braket}
\usepackage{xcolor}
\usepackage{multirow}
\usepackage{xhfill}
\usepackage{array}
\allowdisplaybreaks

\usepackage{graphicx}
\usepackage[colorlinks=true]{hyperref}  
\hypersetup{
    bookmarks=true,         % show bookmarks bar?
    unicode=false,          % non-Latin characters 
    pdftoolbar=true,        % show Acrobat
    pdfmenubar=true,        % show Acrobat 
    pdffitwindow=false,     % window fit to page when opened
    pdfstartview={FitH},    % fits the width of the page to the window
    pdftitle={Orbital currents in insulating and doped antiferromagnets},    % title
    pdfauthor={Mathias Scheurer and Subir Sachdev},     % author
    pdfsubject={},   % subject of the document
    pdfcreator={},   % creator of the document
    pdfproducer={}, % producer of the document
    pdfkeywords={} {} {}, % list of keywords
    pdfnewwindow=true,      % links in new window
    colorlinks=true,       % false: boxed links; true: colored links
    linkcolor=magenta, %red,          % color of internal links (change box color with linkbordercolor)
    citecolor=blue,        % color of links to bibliography
    filecolor=magenta,      % color of file links
    urlcolor=blue           % color of external links
}

\newcommand{\equref}[1]{Eq.~(\ref{#1})}
\newcommand{\equsref}[2]{Eqs.~(\ref{#1}) and (\ref{#2})}
\newcommand{\secref}[1]{Sec.~\ref{#1}}
\newcommand{\figref}[1]{Fig.~\ref{#1}}
\newcommand{\refcite}[1]{Ref.~\onlinecite{#1}}
\newcommand{\refscite}[1]{Refs.~\onlinecite{#1}}

\newcommand{\tableref}[1]{Table~\ref{#1}}
\newcommand{\appref}[1]{Appendix~\ref{#1}}

\newcommand{\CP}{$\mathbb{CP}^1$ }
\newcommand{\CPm}{\mathbb{CP}^1}
\newcommand{\pdagger}{{\phantom{\dagger}}}
\newcommand{\diff}{\mathrm{d}}
\newcommand{\sign}{\,\text{sign}}

\renewcommand{\Re}{\text{Re}}
\renewcommand{\Im}{\text{Im}}
\renewcommand{\vec}[1]{\boldsymbol{#1}}

\newcommand{\beq}{\begin{equation}}
\newcommand{\eeq}{\end{equation}}

\def\bea{\begin{eqnarray}}
\def\eea{\end{eqnarray}}

\definecolor{wrongultramarine}{rgb}{1,0.5,0}

\begin{document}
\title{Orbital currents in insulating and doped antiferromagnets}
\author{Mathias S.~Scheurer}
\affiliation{Department of Physics, Harvard University, Cambridge MA 02138, USA}

\author{Subir Sachdev}
\affiliation{Department of Physics, Harvard University, Cambridge MA 02138, USA}
\affiliation{Perimeter Institute for Theoretical Physics, Waterloo, Ontario, Canada N2L 2Y5}

%===================================================================================================================
\begin{abstract}
We describe square lattice spin liquids which break time-reversal symmetry, while preserving translational symmetry. The states are distinguished by the manner in which they transform under mirror symmetries. All the states have non-zero scalar spin chirality, which implies the appearance of spontaneous orbital charge currents in the bulk (even in the insulator); but in some cases, orbital currents are non-zero only in a formulation with three orbitals per unit cell. The states are formulated using both the bosonic and fermionic spinon approaches. We describe states with $\mathbb{Z}_2$ and U(1) bulk topological order, and the chiral spin liquid with semionic excitations.
The chiral spin liquid has no orbital currents in the one-band formulation, but does have orbital currents in the three-band formulation.
We discuss application to the cuprate superconductors, after postulating that the broken time-reversal and mirror symmetries persist into confining phases which may also break other symmetries. In particular, the broken symmetries of the chiral spin liquid could persist into the N\'eel state.
\end{abstract}
\maketitle

\section{Introduction}

A puzzling feature of the underdoped pseudogap regime of the hole-doped cuprate superconductors is the presence of
numerous experimental signals of time-reversal symmetry breaking and associated orbitals currents \cite{Kapitulnik08,2014PhRvL.112n7001L,2015NatCo...6E7705M,2016arXiv161108603Z,Sonier17,Hayden17}. These features do not appear to be naturally connected to other characteristics of the pseudogap phase: the anti-nodal gap in the fermion spectrum or the symmetry-breaking antiferromagnetic and charge density wave orders observed at lower temperature or doping.

In this paper we shall take the view, following other recent work  \cite{sachdev2017insulators,PhysRevLett.119.227002,ATSS18}, that the orbital currents are characteristics of a parent
`spin liquid' or `topologically ordered' state. 
The topological order induces a gap in the anti-nodal fermion spectrum, and the time-reversal symmetry breaking is an ancilliary feature which is not directly connected to the gap.
The onset of conventional symmetry-breaking orders (such as the N\'eel order) is likely associated with a confinement transition, but we assume that the time-reversal symmetry breaking and the orbital currents survive across such a transition. So this paper will explore the patterns of orbital currents in some reasonable candidate spin liquid states.

In the early studies of square lattice spin liquids, two distinct approaches were used. These represented the spins in terms
of fractionalized spinon operators, using either a canonical Schwinger fermion or Schwinger boson for the spinon operator.
In a more general language, adapted for easier identification of the charged excitations and extension to the doped insulator, these two approaches can be identified with a formalism that transforms to a rotated reference frame in pseudospin or spin space, respectively. These formalisms lead to two distinct SU(2) gauge theories of spin liquids, which we will review in Section~\ref{sec:su2}. 

We find 4 important classes of spin liquids which break time-reversal symmetry, and their broken mirror symmetries are illustrated in Table~\ref{LoopCurrentPattern}. The patterns are shown in both the 
single Cu-orbital model, and the three-orbital CuO$_2$ model, of the cuprate superconductors; these models will be recalled later in this paper.
\begin{table*}[tb]
\begin{center}
\caption{Summary of the four different orbital-current configurations, A, B, C, and D, we consider in this work. In the second and third column, the corresponding loop-current patterns are shown in the one- and three-orbital model, respectively. We focus on one unit cell as the states are all assumed to be invariant under lattice translations. We also indicate (in blue) the generators of the (two-dimensional) residual magnetic point group  consistent with the current patterns. Here, $\Theta$ denotes time-reversal, $\sigma_{x_1x_2}$ reflection at the $x_1x_2$-plane, $\sigma_d$, $\sigma_{d'}$ reflection at the planes going through the diagonals of the square lattice, and $C_{4}$ represents four-fold rotation along the $z$ axis.  The fourth column shows the symmetry properties (using differently textured blue lines) of the overlap (or kinetic energy) $K$ of the orbitals along the different nearest-neighbor bonds in the three-orbital model. As we discuss in detail in \secref{SymmetrySignatures}, some bonds that have identical overlap in the absence of a magnetic field $B_z$ perpendicular to the plane (labeled by the same number $n$ or $n'$, \textit{i.e.}, $K_n=K_{n'}$ at $B_z=0$) will assume different values if a magnetic field is applied along the $z$ direction ($K_n\neq K_{n'}$ for $B_z \neq 0$). This can be used to reveal the non-trivial magnetic symmetries of loop current patterns with experimental probes, such as scanning tunneling microscopy (STM), that are only sensitive to time-reversal-even observables. For $B_z \neq 0$, also the symmetries of the loop currents change, which we have indicated by the red numbers in the third column using the same convention as for $K_n$. The fifth column provides the associated (three-dimensional) magnetic point groups in the absence ($B_z = 0$) and presence ($B_z \neq 0$) of a magnetic field.
Finally, the last column states the respective symmetries of  $e_n$ and $b_n$, defined in \equref{LeadingWilsonLoops} for the three-orbital model, which we use throughout the paper to construct order parameters for the different loop-current patterns.}
\label{LoopCurrentPattern}
 \begin{tabular} {m{1cm}m{3.5cm}m{3.5cm}m{3.5cm}m{2.8cm}m{2.5cm}c} \hline \hline
Pattern & \centering $1$-orbital model & \centering $3$-orbital model  &  \centering Kinetic energies & \centering Mag.~point group & \centering $e_n$, $b_n$ & \\ \hline
\centering A   &  \centering\includegraphics[width=3cm]{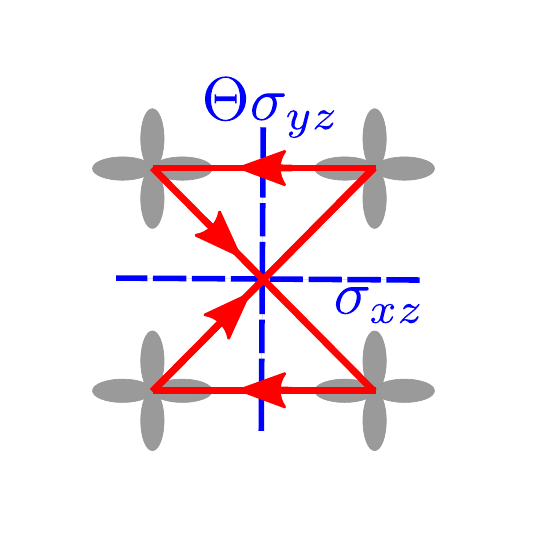} & \centering\includegraphics[width=3cm]{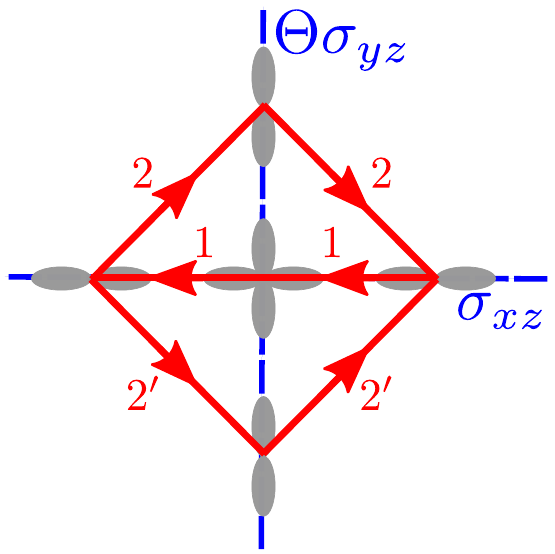} & \centering\includegraphics[width=3cm]{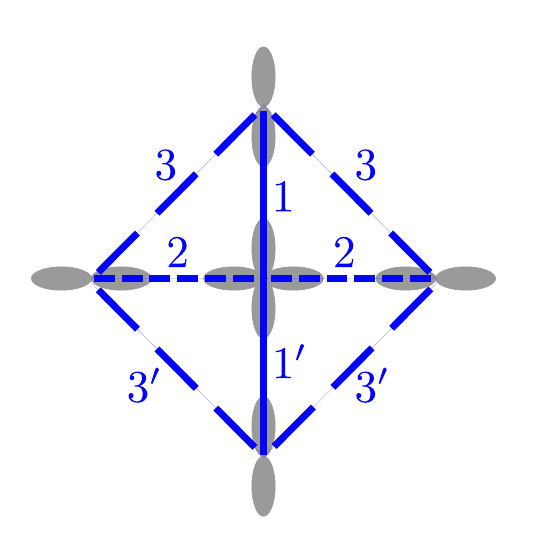}   &  \centering  $m'mm\, (B_z=0)$/ \\  $m'2'm\, (B_z\neq0)$ &  \centering  $b_1=b_2=$ \\ $-b_3=-b_4$, \\ $e_n=e_{n+1}$  & \\ \hline 
\centering B  &  \centering\includegraphics[width=3cm]{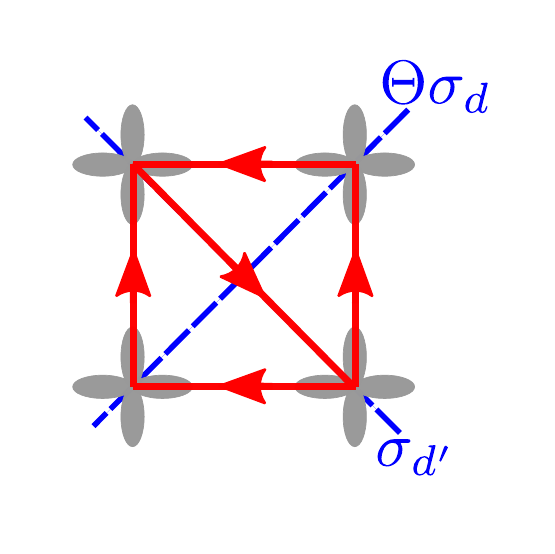} & \centering\includegraphics[width=3cm]{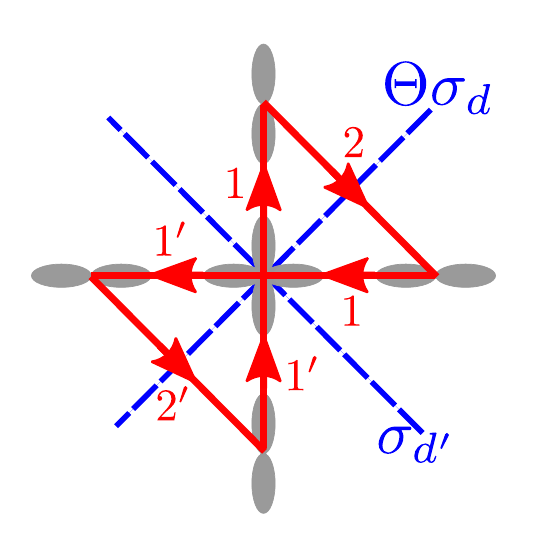} & \centering \includegraphics[width=3cm]{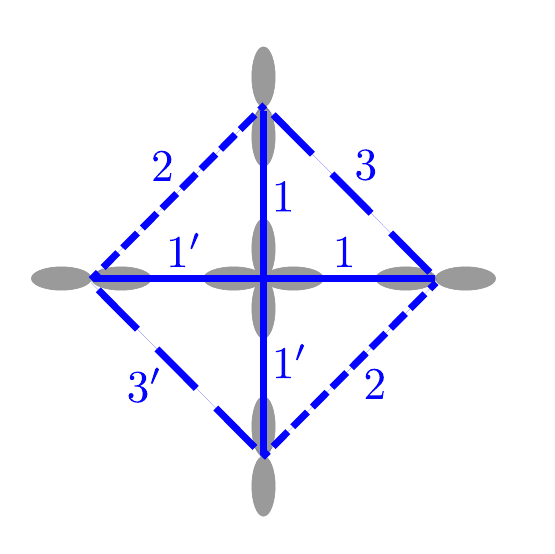} & \centering $m'mm\, (B_z=0)$/ \\  $m'2'm\, (B_z\neq0)$ & \centering $b_1=-b_3$, $b_2=b_4=0$, $e_n=e_{n+2}$  & \\ \hline  \centering C  &  \centering All orbital\\ currents vanish & \centering\includegraphics[width=3cm]{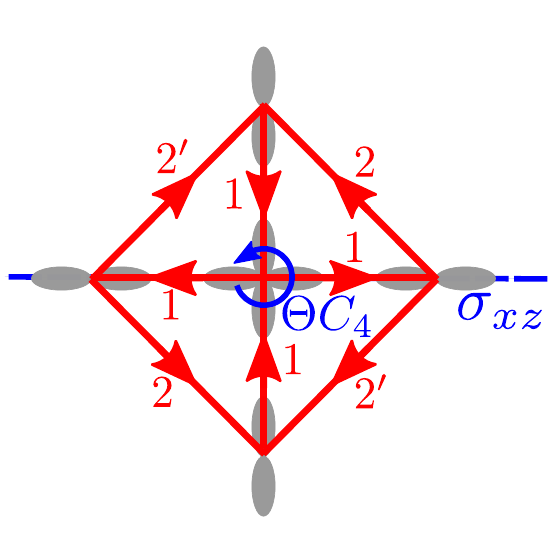} & \centering \includegraphics[width=3cm]{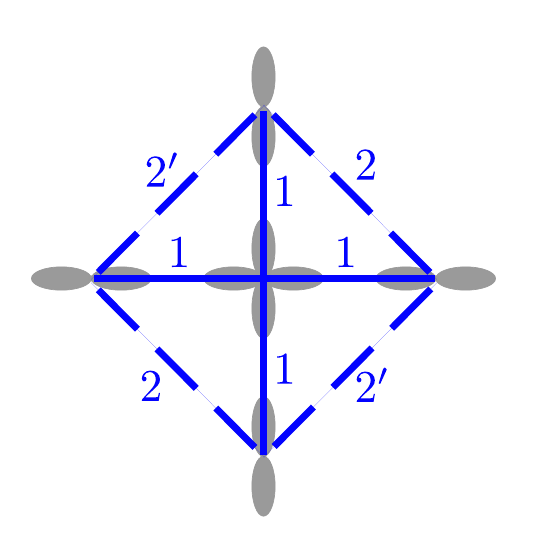} & \centering $\frac{4'}{m}mm'\, (B_z=0)$/ \\  $m'm'm\, (B_z\neq0)$  & \centering $b_1=b_3=$ \\$-b_2=-b_4$, \\ $e_n=e_{n+1}$  & \\\hline
\centering D  &  \centering All orbital\\ currents vanish & \centering\includegraphics[width=3cm]{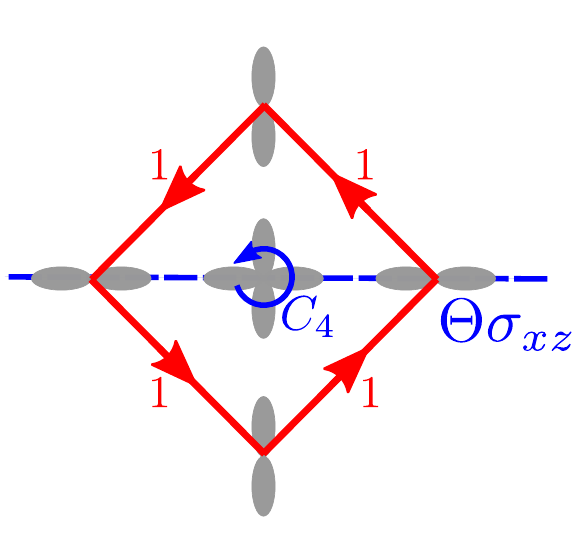} & \centering \includegraphics[width=3cm]{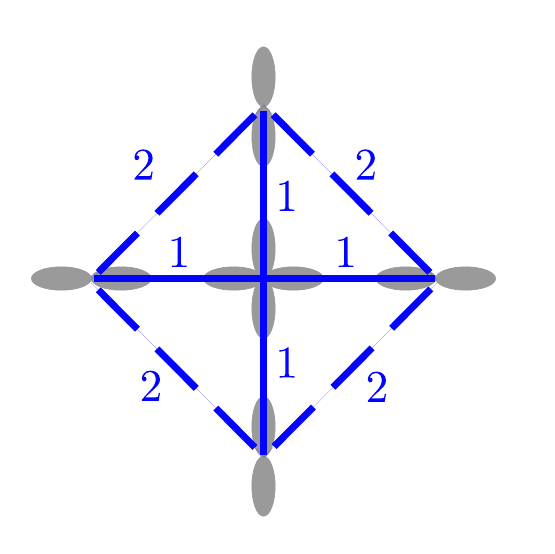} & \centering $\frac{4}{m}m'm'\, (B_z=0)$/ \\  $\frac{4}{m}m'm'\, (B_z\neq0)$ & \centering $b_n=b_{n+1}$, $e_n=e_{n+1}$ & \\ \hline\hline
 \end{tabular}
\end{center}
\end{table*}
A complementary study of broken time-reversal and mirror symmetries in weakly-coupled Fermi liquids was presented by Sun and Fradkin \cite{SunFradkin08}, and there are some connections to our classifications. We note the relationships between the states in Table~\ref{LoopCurrentPattern} and previous studies:
\begin{itemize}
\item 
Patterns B and C were proposed by Varma \cite{SimonVarma,VarmaPRB2006} in the three-orbital models of the cuprate superconductors which do not have fractionalization. Pattern B was also discussed by Sun and Fradkin \cite{SunFradkin08} in a weak-coupling theory.
\item
Patterns A and B appeared in studies of $\mathbb{Z}_2$ spin liquids using bosonic \cite{sachdev2017insulators,PhysRevLett.119.227002} and fermionic \cite{ATSS18} spinons, both in the 1-orbital model. 
\item Pattern D appeared in the study by Sun and Fradkin \cite{SunFradkin08} and others \cite{HeMoore12,HeLee14}.
\end{itemize}
%We note that pattern D transforms as the orbital magnetic field perpendicular to the CuO$_2$ planes under all symmetries of the system; it is pattern D alone that does not possess a mirror plane symmetry (without composing with time-reversal) along any orientation within the plane of the square lattice. Consequently, it is the only pattern which yields a non-zero Kerr effect, and a non-zero anomalous Hall effect in the metallic case.

We note that among these 4 patterns, it is pattern D alone which does not possess a mirror plane symmetry (without composing with time-reversal) along any orientation within the plane of the square lattice (it has the same symmetries as the orbital magnetic field perpendicular to the CuO$_2$ planes). Consequently, pattern D is the only pattern which yields a non-zero Kerr effect, and a non-zero anomalous Hall effect in the metallic case.

In this paper, we will provide a bosonic spinon theory of patterns A, B, and C in the three-orbital model in Section~\ref{BosonicSpinons}. As is noted in Table~\ref{LoopCurrentPattern}, the orbital currents in pattern C are non-zero only in the three-orbital model.

We will also present both bosonic and fermionic spinon theories of pattern D. In the bosonic spinon approach, described in Section~\ref{BosonicSpinons}, the saddle-point yields a state with $\mathbb{Z}_2$ ({\it i.e.\/} like in the toric code) or U(1) topological order.
In contrast, the fermionic spinon approach, described in Section~\ref{FermionicSpinons}, leads to an induced Chern-Simons term for the emergent gauge field, and so the resulting state is a chiral spin liquid \cite{KL87,WWZ89}. 
So for pattern D, we obtain distinct spin liquid states from the bosonic and fermionic spinon approaches.
Orbital charge currents vanished in previous single-band studies of the chiral spin liquid. Here, we will present a three-band formulation of the chiral spin liquid, and show that it has spontaneous orbital currents as in pattern D.

We will also discuss additional measurable quantities that allow to distinguish between the four different patterns A-D, both for the one- and three-orbital model. For instance, all four patterns exhibit distinct combinations (as dictated by the magnetic point symmetries) of scalar spin chiralities $\braket{\hat{\vec{S}}_i \cdot (\hat{\vec{S}}_j \times \hat{\vec{S}}_k)}$, where $\hat{\vec{S}}_i$ is the spin operator on site $i$ and $i,j,k$ are nearest neighbors. We also show that the different patterns can be detected in time-reversal invariant observables, such as the overlap of the Wannier states along the different bonds which is, in principle, accessible by STM measurements: as summarized in \tableref{LoopCurrentPattern}, the combination of the magnetic point symmetries of the different patterns and the symmetry reduction when a magnetic field is applied lead to a unique deformation of the bonds within the unit cell that can be used to distinguish experimentally between the four different states.

We now outline the remainder of the paper. We will begin in \secref{sec:su2} by recalling basic aspects of the bosonic and fermionic spinon theories of spin liquids. We will do this in a unified formalism which clarifies the relationship between the two approaches. 
Section~\ref{BosonicSpinons} presents lattice mean-field theories of the bosonic spinon theory of all 4 patterns of Table~\ref{LoopCurrentPattern}; we will start with the three-orbital model in \secref{ThreeOrbiModLatt}, that allows for non-zero loop currents for all patterns, and then discuss how states with the same symmetries as those of the four loop-current patterns can be realized in the
one-orbital model (see \secref{OneOrbitalModel}). While we mainly focus on the charge degrees of freedom in these two subsections, we discuss the spin sector of the theory in \secref{CP1TheoryDescription}. 
The discussion of the complementary fermionic spinon approach can be found \secref{FermionicSpinons}, where we focus on pattern D. 
We analyze the symmetry signatures in the presence of a magnetic field of the different patterns in \secref{SymmetrySignatures} and comment on experimental consequences.
Finally, \secref{Conclusion} summarizes our results and discusses application to the cuprates.

\section{SU(2) gauge theories}
\label{sec:su2}

It is useful to first express the electron operator in a formalism that treats spin and Nambu pseudospin rotations at an equal footing.
So we write the electron operator $c_{i \alpha}$ ($\alpha = \uparrow, \downarrow$ is the spin index) in the matrix form
\beq
C_i = \left(
\begin{array}{cc}
c_{i \uparrow} & - c_{i \downarrow}^\dagger \\
c_{i \downarrow} & c_{i \uparrow}^\dagger
\end{array}
\right) \label{Xc}
\eeq
This matrix obeys the relation
\beq
C_i^\dagger = \sigma^y C_i^T \sigma^y.
\eeq
Global SU(2)$_s$ spin rotations, $U_s$ act on $C$ by left multiplication
while global SU(2)$_c$ Nambu pseudospin rotations, $U_c$ act on $C$ by right multiplication.
The electron spin operator ${\bm S}_i$ is given by
\beq
{\bm S}_i = \frac{1}{4} \mbox{Tr} ( C_i^\dagger {\bm \sigma} C_i )\,,
\eeq
and the electron Nambu pseudospin operator ${\bm T}_i$ is given by
\bea
{\bm T}_i &=& \frac{1}{4} \mbox{Tr} ( C_i^\dagger C_i  {\bm \sigma} ) \nonumber \\
&=& 
\frac{1}{2} \left(
c_{i\downarrow}^\dagger c_{i \uparrow}^\dagger + c_{i\uparrow} c_{i \downarrow} ,
i\left(c_{i\downarrow}^\dagger c_{i \uparrow}^\dagger - c_{i\uparrow} c_{i \downarrow} \right), \right. \nonumber \\
&~&~~~~~~~~~\left.
c_{i\uparrow}^\dagger c_{i \uparrow} + c_{i\downarrow}^\dagger c_{i \downarrow} -1 \right)\,.
\eea
Note that $T_z$ is just the electronic charge operator. This implies that a generic chemical potential, $\mu$, in the Hamiltonian will break global SU(2)$_c$ pseudospin symmetry down to the charge U(1)$_c$.
All the ${\bm S}_i$ commute with all the ${\bm T}_i$.
 
It is also useful to note that the squares of the spin and pseudospin operators are
\bea
{\bm S}_i^2 &=&- \frac{3}{2} \left( c_{i \uparrow}^\dagger c_{i \uparrow}^\pdagger - \frac{1}{2} \right)  \left( c_{i \downarrow}^\dagger c_{i \downarrow}^\pdagger - \frac{1}{2} \right) + \frac{3}{8} \nonumber \\
{\bm T}_i^2 &=& \frac{3}{2} \left( c_{i \uparrow}^\dagger c_{i \uparrow}^\pdagger - \frac{1}{2} \right)  \left( c_{i \downarrow}^\dagger c_{i \downarrow}^\pdagger - \frac{1}{2} \right) + \frac{3}{8}
\eea
So the Hubbard interaction, $U$, can be written either in terms of ${\bm S}_i^2$ or ${\bm T}_i^2$, and the chemical potential preserves global SU(2)$_c$ pseudospin symmetry only at $\mu = U/2$.

\subsection{Bosonic spinons}
\label{sec:bosons}

The bosonic spinon formulation is obtained by transforming to the rotating reference frame in spin space. 
We introduce a SU(2) matrix $R_{si}$ which transforms the SU(2)$_s$ index
to a rotating reference frame by defining \cite{PhysRevB.80.155129,PhysRevLett.119.227002,Scheurer201720580}
\beq
C_i = R_{si} \Psi_i \label{RPsi}
\eeq
where the $\Psi$ are fermionic chargons defined as in Eq.~(\ref{Xc})
\beq
\Psi_i = \left(
\begin{array}{cc}
\psi_{i +} & - \psi_{i -}^\dagger \\
\psi_{i - } & \psi_{i +}^\dagger
\end{array}
\right) \label{Psipsi}
\eeq
while $R_{si}$ is a c-number SU(2) matrix
\beq
R_{si} = \left(
\begin{array}{cc}
z_{i \uparrow} & - z_{i \downarrow}^\ast \\
z_{i \downarrow} & z_{i \uparrow}^\ast
\end{array}
\right) \label{Rz}
\eeq
with
\beq
|z_{i\uparrow}|^2 + |z_{i\downarrow}|^2 = 1
\eeq
so that $R_{si}^\pdagger R_{si}^\dagger = R_{si}^\dagger R_{si}^\pdagger = \mathds{1}$. Note that the $z_{i \alpha}$ are not canonical bosons in this formalism. At low energies, the $z_{i \alpha}$ ultimately map onto degrees of freedom usually obtained via the Schwinger boson formulation \cite{PhysRevLett.119.227002,Scheurer201720580}.
But the intermediate steps in the present mapping are different from those starting with canonical Schwinger bosons \cite{NRSS90}.
Only the fermionic chargons are canonical in our formalism here.

The parameterization in Eq.~(\ref{RPsi}) introduces a SU(2) gauge invariance which we will denote as SU(2)$_{sg}$.
The transformations of the fields introduced under the various SU(2)s are:
\bea
\text{SU(2)}_{sg} &:& \quad C \rightarrow C, \quad \Psi \rightarrow  U_{sg} \Psi , \quad R_s \rightarrow R_s U^\dagger_{sg} \nonumber \\
\text{SU(2)}_s &:& \quad C \rightarrow U_s C, \quad \Psi \rightarrow  \Psi , \quad R_s \rightarrow U_s R_s \nonumber \\
\text{SU(2)}_c &:& \quad C \rightarrow C U_c , \quad \Psi \rightarrow \Psi U_c , \quad R_s \rightarrow R_s  \label{SG}
\eea
The $R_s$ transform under global spin rotations, and hence they carry spin: so the $R_s$
are the bosonic spinons. We chose the parameterization in Eq.~(\ref{Rz}) so that $(z_\uparrow, z_\downarrow)$ transforms as a doublet under 
SU(2)$_s$.

In this spin rotating frame, the electron pseudospin operator is expressed only in terms of $\Psi$
\beq
{\bm T}_i = \frac{1}{4} \mbox{Tr} ( \Psi_i^\dagger \Psi_i {\bm \sigma})\,,
\eeq
but the spin does not factorize
\beq
{\bm S}_i = \frac{1}{4} \mbox{Tr} (\Psi_i^\dagger R_{si}^\dagger {\bm \sigma} R_{si} \Psi_i)\,.
\eeq
The expressions for the squares of the spin and pseudospin are independent of $R_{si}$, and parallel those for the electrons
\bea
{\bm S}_i^2 &=&- \frac{3}{2} \left( \psi_{i +}^\dagger \psi_{i +}^\pdagger - \frac{1}{2} \right)  \left( \psi_{i -}^\dagger \psi_{i -}^\pdagger - \frac{1}{2} \right) + \frac{3}{8} \nonumber \\
{\bm T}_i^2 &=& \frac{3}{2} \left( \psi_{i +}^\dagger \psi_{i +}^\pdagger - \frac{1}{2} \right)  \left( \psi_{i -}^\dagger \psi_{i -}^\pdagger - \frac{1}{2} \right) + \frac{3}{8}
\eea

\subsection{Fermionic spinons}
\label{sec:fermions}

The fermionic spinon formulation is obtained by transforming to the rotating reference frame in pseudospin space. The analysis parallels that in Section~\ref{sec:bosons}, with the spin and pseudospin exchanging roles. 
Now we introduce a SU(2) matrix $R_{ci}$ which transforms the SU(2)$_c$ index
to a rotating reference frame by defining \cite{LeeWenRMP,Wen2,XS10}
\beq
C_i = F_i R_{ci} \label{FR}
\eeq
where the $F$ are fermions (`spinons') defined as in Eq.~(\ref{Xc})
\beq
F_i = \left(
\begin{array}{cc}
f_{i \uparrow} & - f_{i \downarrow}^\dagger \\
f_{i \downarrow} & f_{i \uparrow}^\dagger
\end{array}
\right) \label{Ff}
\eeq
while $R_{ci}$ is a c-number SU(2) matrix
\beq
R_{ci} = \left(
\begin{array}{cc}
b_{i 1} & b_{i 2} \\
-b_{i 2}^\ast & b_{i 1}^\ast
\end{array}
\right) \label{Rb}
\eeq
with
\beq
|b_{i1}|^2 + |b_{i2}|^2 = 1
\eeq
so that $R_{ci}^\pdagger R_{ci}^\dagger = R_{ci}^\dagger R_{ci}^\pdagger = \mathds{1}$.
Note that the $b$'s are not canonical Bose operators, unlike the approach in Ref.~\onlinecite{LeeWenRMP}.
Instead we have introduced them as components of a $c$-number SU(2) matrix; in the path integral formulation, the $b$'s
will acquire dynamics only after the fermions have been integrated out. 

The parameterization in Eq.~(\ref{FR}) introduces a SU(2) gauge invariance which we will denote as SU(2)$_{cg}$.
The transformations of the fields introduced under the various SU(2)s are:
\bea
\text{SU(2)}_{cg} &:& \quad C \rightarrow C, \quad F \rightarrow  F U_{cg} , \quad R_c \rightarrow U^\dagger_{cg} R_c \nonumber \\ 
\text{SU(2)}_s &:& \quad C \rightarrow U_s C, \quad F \rightarrow U_s F , \quad R_c \rightarrow R_c \nonumber \\
\text{SU(2)}_c &:& \quad C \rightarrow C U_c , \quad F \rightarrow F , \quad R_c \rightarrow R_c U_c 
\eea
Note that the $R_c$ transform under global pseudospin rotations, and hence they carry the electronic charge: so the $R_c$
are the bosonic `chargons'. We chose the parameterization in Eq.~(\ref{Rb}) so that $(b_1, b_2)$ transforms as a doublet under 
SU(2)$_c$. In our formulation, it is crucial to distinguish between SU(2)$_{c}$ and SU(2)$_{cg}$; in contrast, the setup of Ref.~\onlinecite{LeeWenRMP}
did not make this distinction, and only considered SU(2)$_{cg}$

In this pseudospin rotating frame, the electron spin operator is expressed only in terms of $F$
\beq
{\bm S}_i = \frac{1}{4} \mbox{Tr} ( F_i^\dagger {\bm \sigma} F_i )\,,
\eeq
but the pseudospin does not factorize
\beq
{\bm T}_i = \frac{1}{4} \mbox{Tr} ( R_{ci}^\dagger F_i^\dagger  F_i  R_{ci} {\bm \sigma} )\,.
\eeq
The expressions for the squares of the spin and pseudospin are independent of $R_{ci}$, and parallel those for the electrons
\bea
{\bm S}_i^2 &=&- \frac{3}{2} \left( f_{i \uparrow}^\dagger f_{i \uparrow}^\pdagger - \frac{1}{2} \right)  \left( f_{i \downarrow}^\dagger f_{i \downarrow}^\pdagger - \frac{1}{2} \right) + \frac{3}{8} \nonumber \\
{\bm T}_i^2 &=& \frac{3}{2} \left( f_{i \uparrow}^\dagger f_{i \uparrow}^\pdagger - \frac{1}{2} \right)  \left( f_{i \downarrow}^\dagger f_{i \downarrow}^\pdagger - \frac{1}{2} \right) + \frac{3}{8}
\eea

%=================================================================================================================== 
\section{Bosonic spinons and fermionic chargons}\label{BosonicSpinons}
In this section, we assume that the charge degrees of freedom have fermionic statistics, while the spin degrees of freedom have bosonic statistics.
We provide descriptions in the three-orbital model in \secref{ThreeOrbiModLatt}, in the one-orbital model in  \secref{OneOrbitalModel}, and using the \CP theory of fluctuating antiferromagnetism in \secref{CP1TheoryDescription}.

\subsection{Three-orbital model}
\label{ThreeOrbiModLatt}
\begin{figure}[tb]
\begin{center}
\includegraphics[width=\linewidth]{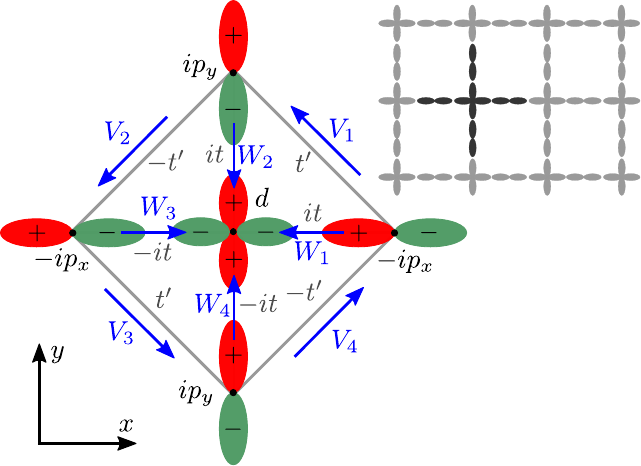}
\caption{CuO$_2$ lattice (upper right) and illustration of the conventions used in this paper shown in one unit cell of the three-orbital model (lower left). The non-zero hopping matrix elements $t_{ij}$ on the different bonds, see \equref{Emery}, and the gauge connections of the chargons, see \equref{EffectiveChargonHam}, are indicated in black and blue, respectively.}
\label{PhaseConvetions}
\end{center}
\end{figure}

As it allows for finite loop-currents for all four patterns in \tableref{LoopCurrentPattern}, let us begin with the three-orbital model \cite{EmeryModel,VARMA1987681} of the CuO$_2$ planes. %For completeness and to illustrate the difficulties, we will come back to the one-orbital model in \secref{OneOrbitalModel}.  
The Hamiltonian $H=H_0+H_{\text{int}}$ of the three-orbital model consists of two parts. The noninteracting part, $H_0$, can be written as 
\begin{subequations}\begin{align}\begin{split}
H_0 =  &-t {\sum_{j,\alpha}}^\prime\sum_{s=\pm}\sum_{l=x,y} s \left(i\, d^\dagger_{j,\alpha}p^\pdagger_{l,j+s\frac{\vec{e}_l}{2},\alpha}  + \text{H.c.}\right) \\
  &-t' {\sum_{j,\alpha}}^\prime\sum_{s,s'=\pm} ss' \left(p^\dagger_{y,j+s\frac{\vec{e}_y}{2},\alpha}p^\pdagger_{x,j+s'\frac{\vec{e}_x}{2},\alpha}  + \text{H.c.}\right) \\ 
  &-\Delta {\sum_{j}}^\prime  n_{p,j} - \mu {\sum_{j}}^\prime  n_{j},   \label{Emery}
\end{split}\end{align}
which is also summarized graphically in \figref{PhaseConvetions}.
Here, $d_{j,\alpha}$ and $p_{x,j,\alpha}$ ($p_{y,j,\alpha}$) describe the annihilation of an electron in the Cu-$d$ and in the O-$p_x$ (O-$p_y$) orbital of spin $\alpha$ and on site $j$ of the CuO$_2$ plane. We include both nearest-neighbor hopping $t$ between the Cu-$d$ and the O-$p$ orbitals as well as the diagonal hopping matrix elements $t'$ between O-$p_x$ and O-$p_y$. The total density of electrons and of those residing on oxygen atoms only are denoted by $n_{j}=\sum_\alpha d^\dagger_{j,\alpha}d^\pdagger_{j,\alpha}+n_{p,j}$ and $n_{p,j}=\sum_{s,l} p^\dagger_{l,j+s\frac{\vec{e}_l}{2},\alpha}p^\pdagger_{l,j+s'\frac{\vec{e}_l}{2},\alpha} $, respectively. The prefactors $\mu$ and $\Delta$ multiplying these operators in the Hamiltonian (\ref{Emery}) are the chemical potential and the energetic on-site splitting between the Cu and O atoms. In \equref{Emery}, the primes on the summation symbols indicate that the sum only involves sites $j\in\mathbb{Z}^2$ of the square lattice of the Cu atoms. 
 
To keep the notation as compact as possible we will use, from now on, the following equivalent representation of the noninteracting part of the Hamiltonian:
\begin{equation}
H_0 = - \sum_{ij}\sum_{\alpha} t_{ij} c^\dagger_{i\alpha}c^\pdagger_{j\alpha} + \sum_{j}\sum_{\alpha} \left(\Delta_j -\mu \right)c^\dagger_{j\alpha}c^\pdagger_{j\alpha},
\end{equation}\label{EmeryGeneral}\end{subequations}
where the summations run over all Cu and O sites of the CuO$_2$ layers of the system and $c_{j\alpha} = d_{j,\alpha}$, or $c_{j\alpha} = p_{x,j,\alpha}$, $c_{j\alpha} = p_{y,j,\alpha}$ depending on whether $j$ refers to a site with Cu-$d$ (integer $j_x$, $j_y$), or O-${p_x}$, O-${p_x}$ orbital (half-integer $j_x$ or $j_y$) as relevant low-energy degree of freedom.  Accordingly, $\Delta_j\in\{-\Delta,0\}$ and  $t_{ij}\in\{\pm i t,\pm t'\}$ with $t^*_{ij}=t_{ji}$ as required by Hermiticity.

We take the interaction $H_{\text{int}}$ to be of the spin-fermion form 
\begin{equation}
H_{\text{int}} = - g {\sum_{j}}^{\prime}\sum_{\alpha\beta}  c^\dagger_{j\alpha} \vec{\sigma}_{\alpha\beta} c^\pdagger_{j\beta} \cdot \vec{\Phi}_{j} + H_\Phi, \label{SpinFermionCoupling}
\end{equation}
where $\vec{\sigma}=(\sigma_x,\sigma_y,\sigma_z)^T$ is the vector of Pauli matrices and $\vec{\Phi}_{j}$ a bosonic field describing collective spin-fluctuations. For a complete description, we also need to add a contribution $H_\Phi$ to the Hamiltonian that determines the dynamics of the collective mode $\vec{\Phi}_{j}$. However, for the following analysis, we will not have to specify $H_\Phi$ explicitly but, instead and more generally, assume an appropriately chosen form of $H_\Phi$ to tune between the different phases that we discuss below. 

For simplicity, we assume that the spin-fluctuations only couple to the electrons in the Cu-$d$ orbitals and neglect any coupling to electrons residing on the oxygen atoms; this is already indicated by the prime in the sum in \equref{SpinFermionCoupling}. We will see below that it already allows for all relevant loop-current-order phases. The generalization to also including couplings to the oxygen atoms is straightforward but does not provide additional crucial physical insights for our analysis.

\subsubsection{Effective Hamiltonian of the chargons}
To be able to describe a pseudogap phase in the absence of broken translational symmetry, we follow previous work \cite{PhysRevB.80.155129,sachdev2017insulators,PhysRevLett.119.227002,Scheurer201720580} and transform to a rotating reference frame in spin space, \textit{i.e.}, use the description introduced in \secref{sec:bosons}. This means that we rewrite the electronic operator $c_{j\alpha}$ according to \equref{RPsi} or, more compactly,
\begin{equation}
c_{j\alpha} = \sum_{\beta=\pm}\left(R_{sj}\right)_{\alpha \beta} \psi_{j\beta},  \quad R_{sj} \in \text{SU}(2), \label{Trafo}
\end{equation}
which will allow us to conveniently describe phases with topological order. Physically, \equref{Trafo} corresponds to the fractionalization of the electron's spin, carried by the ``spinon'' $R_{sj}$, and its charge degree of freedom, carried by the ``chargon'' $\psi_{j}=(\psi_{j+},\psi_{j,-})^T$. It also automatically introduces the local $\text{SU}(2)_{sg}$ gauge symmetry, defined in \equref{SG}, which reads in spinor notation as
\begin{equation}
 \psi_{j}  \rightarrow U_{sgj}\psi_{j}, \quad R_{sj}  \rightarrow R_{sj}U^\dagger_{sgj}, \quad U_{sgj} \in \text{SU}(2). \label{SUGauge}
 \end{equation} 
Inserting the transformation (\ref{Trafo}) into the spin-fermion coupling (\ref{SpinFermionCoupling}), we obtain the chargon-Higgs coupling
\begin{equation}
H_{\text{int}} = - g {\sum_{j}}^{\prime}\sum_{\alpha\beta}  \psi^\dagger_{j\alpha} \vec{\sigma}_{\alpha\beta} \psi^\pdagger_{j\beta} \cdot \vec{H}_{j} + H_H, \label{ChargonHiggsCoupling}
\end{equation}
with Higgs field $\vec{H}_{j}$ defined according to
\begin{equation}
    \vec{\sigma}\cdot\vec{H}_{j} = R^\dagger_{sj} \vec{\sigma} R^\pdagger_{sj} \cdot \vec{\Phi}_{j}. \label{HiggsOrderParamRelation}
\end{equation}
From this definition, we directly see that the Higgs field transforms under the adjoint representation of $\text{SU}(2)_{sg}$. Exactly as its parent term $H_\Phi$ in \equref{SpinFermionCoupling}, $H_H$ will not be specified explicitly here. 
Note that, due to our approximation of neglecting the spin-fermion interaction on the oxygen sites, also the Higgs field only couples to the chargons residing on the Cu sites.
For notational simplicity, we will absorb the coupling constant $g$ into the definition of the Higgs field.

Since we are mainly interested in loop-current patterns, we will first focus on the chargons. In \secref{CP1TheoryDescription}, we will come back to the spin degrees of freedom and derive the corresponding \CP theory. The effective Hamiltonian $H_\psi$ of the chargons is obtained by inserting the transformation (\ref{Trafo}) into \equref{EmeryGeneral} and performing a decoupling of the quartic spinon-chargon terms. Adding the coupling (\ref{ChargonHiggsCoupling}) to the Higgs field, it reads as 
\begin{align}\begin{split}
&H_\psi = - \sum_{ij}\sum_{\alpha\beta} t_{ij} \psi^\dagger_{i\alpha} \left(U_{ij}\right)_{\alpha\beta} \psi^\pdagger_{j\beta}  \\ &+ \sum_{j}\sum_{\alpha} \left(\Delta_j -\mu \right)\psi^\dagger_{j\alpha}\psi^\pdagger_{j\alpha} - {\sum_{j}}^{\prime}\sum_{\alpha\beta}  \psi^\dagger_{j\alpha} \vec{\sigma}_{\alpha\beta} \psi^\pdagger_{j\beta} \cdot \vec{H}_{j} \end{split} \label{EffectiveChargonHam}
\end{align}
with unitary $U_{ij}$ satisfying $U_{ij} = \left(U_{ji}\right)^\dagger$ as required by Hermiticity and representing the $\text{SU}(2)$ gauge connection felt by the chargons. Note that there are in general also renormalizations of the magnitude of the hopping matrix elements, $t_{ij} \rightarrow Z_{ij} t_{ij}$, for the chargons as has been explicitly demonstrated in \refcite{Scheurer201720580}. As it is not relevant to our discussion here and for notational simplicity, we will neglect the renormalization factors $Z_{ij}$ in this work.

There are configurations of the gauge connections $U_{ij}$ and of the Higgs field that lead to loop currents. This will be shown explicitly below, by evaluation of the expectation value of the operator 
\begin{equation}
J_{ij} = -J_{ji} = - i t_{ij} \sum_{\alpha\beta} \psi^\dagger_{i\alpha} \left(U_{ij}\right)_{\alpha\beta} \psi^\pdagger_{j\beta} + \text{H.c.} \label{FormOfCurOp}
\end{equation}
for the current from site $i$ to site $j$ in the ground state of the chargon Hamiltonian $H_\psi$. The form (\ref{FormOfCurOp}) of $J_{ij}$ follows formally from the continuity equation (we set the electron charge to be $-1$),
\begin{equation}
\partial_t Q_i =  -\sum_j J_{ji}, \qquad Q_i = -\sum_\alpha \psi^\dagger_{i\alpha} \psi^\pdagger_{i\alpha}.
\end{equation}

\subsubsection{Symmetry analysis and loop currents}
\label{SymmetryAnalysisCurrents}
To organize the search for possible ans\"atze for $\vec{H}_j$ and $U_{ij}$ that yield loop-current patterns, we first analyze the time-reversal and space-group symmetries of the chargon Hamiltonian for given $\vec{H}_j$ and $U_{ij}$. 

Naively, one would expect that the mean-field Hamiltonian (\ref{EffectiveChargonHam}) preserves a space group symmetry $g$ if and only if it is explicitly invariant under $g$, \textit{i.e.}, invariant under replacing $\psi_j \rightarrow \psi_{gj}$. Recalling the gauge invariance (\ref{SUGauge}) associated with rewriting the electronic operator in terms of spinons and chargons, we see that this is only sufficient but not necessary; instead, the mean-field Hamiltonian $H_\psi$ respects the symmetry $g$ if and only if there is $G_g(j)\in\text{SU}(2)$ such that the Hamiltonian is invariant under
 \begin{equation}
\psi_j \longrightarrow G_g(j)\psi_{gj}. \label{SpaceGroupTrafo}
 \end{equation}
Similarly, time-reversal symmetry is preserved if the Hamiltonian commutes with the antiunitary operator $\Theta$ defined via
\begin{equation}
 \Theta \psi_{j}\Theta^\dagger = i\sigma_y G_\Theta(j) \psi_{j} \label{DefinitionTRS}
\end{equation}
for some properly chosen $G_\Theta(j)\in\text{SU}(2)$. Note that the additional matrix $i\sigma_y$ appearing in \equref{DefinitionTRS} is just a matter of choice as it could have, equally well, been absorbed into $G_\Theta(j)$. %However, it will turn out to be convenient below (see \secref{PSGAnalysis}).

One important immediate consequence of \equref{DefinitionTRS} is that $\vec{H}_i\neq 0$ is required to have loop currents: Applying the transformation (\ref{DefinitionTRS}) with $G_\Theta(j) = \sigma_0$ to the chargon Hamiltonian (\ref{EffectiveChargonHam}), the Ansatz transforms as $U_{ij} \rightarrow \sigma_y U_{ij}^* \sigma_y = U_{ij}$ (due to unitarity of $U_{ij}$) and  $\vec{H}_i \rightarrow  -\vec{H}_i$. Consequently, only if $\vec{H}_i \neq 0$, time-reversal can be broken which, in turn, is necessary to have non-vanishing currents in the system.

\renewcommand{\arraystretch}{1.4}
\begin{table}[bt]
\begin{center}
\caption{Summary of the transformation behavior of $e_n$ and $b_n$ defined in \equref{LeadingWilsonLoops} under the point group operations, $C_4$, $\sigma_{yz}$, and time-reversal $\Theta$. As in \tableref{LoopCurrentPattern}, $C_4$ and $\sigma_{yz}$ denote $\pi/4$ rotation along the $z$ axis and reflection at the $yz$ plane, respectively. As in the main text, we identify $n=5\equiv 1$ cyclically.}
\label{TransformOfeb}
 \begin{tabular}{ccc} \hline \hline
 \phantom{mmmm} & \phantom{mmmmee'} $e_n$ \phantom{mmmmee'} & \phantom{mmmmee'} $b_n$ \phantom{mmmmee'}   \\ \hline
   $C_4$  & $e_n \rightarrow e_{n+1}$ & $b_n \rightarrow b_{n+1}$   \\
$\sigma_{yz}$  & $\,\,e_1 \leftrightarrow e_{2}$, $e_3 \leftrightarrow e_{4}\,\,$ & $\,\,b_1 \leftrightarrow -b_{2}$, $b_3 \leftrightarrow -b_{4}\,\,$   \\ 
$\Theta$ & $e_n \rightarrow e_n$  & $b_n \rightarrow -b_n$    \\\hline \hline
 \end{tabular}
\end{center}
\end{table}

Instead of only performing a symmetry analysis based on \equsref{SpaceGroupTrafo}{DefinitionTRS}, we will also consider gauge-invariant quantities, denoted by $e_n$ and $b_n$ in the following, with non-trivial transformation behavior to check and illustrate the discussion. This will also allow for a direct connection between $e_n$, $b_n$ and the loop currents.
Let us use $W_n(j)$ and $V_n(j)$, $n=1,2,3,4$, to denote the $\text{SU}(2)$-gauge connections $U$ corresponding to the four Cu-O and O-O bonds in the unit cell (with Cu atom at site $j$) as shown in \figref{PhaseConvetions}. We define   
\begin{subequations}\begin{align}
e_n(j) &= \text{Tr}\left[W_{n}^\dagger(j) V^\pdagger_n(j) W^\pdagger_{n+1}(j)\right],  \\
b_n(j) &= i H_j^a \,\text{Tr}\left[\sigma_a W_{n}^\dagger(j) V^\pdagger_n(j) W^\pdagger_{n+1}(j)\right],   \label{DefinitionOfbs}
\end{align}\label{LeadingWilsonLoops}\end{subequations}
where $n=1,2,3,4$ and we have made the identifications $V_{5}\equiv V_{1}$ and $W_{5}\equiv W_{1}$ to keep the notation compact. 
Taking advantage of the fact that the unitarity of $V_n$ and $W_n$ implies $\sigma_y V_n \sigma_y = V^*_n$ and similarly for $W_n$, it is easy to see that both $e_n(j)$ and $b_n(j)$ are Hermitian. From \equref{DefinitionTRS}, we see that $e_n$ and $b_n$ are even and odd under time-reversal, respectively, while the behavior under spatial symmetries follows from \equref{SpaceGroupTrafo}. The transformation behavior of $e_n(j)$ and $b_n(j)$ is summarized in \tableref{TransformOfeb}. Note further that $e_n(j)$ ($b_n(j)$) is even (odd) under reversal of the direction of the loop or, in other words, $e_1(j) \rightarrow e_1(j)$ ($b_1(j) \rightarrow -b_1(j)$) under reflection $\sigma_d$ at the plane along $x=y$, parallel to the $z$ axis and going through the Cu atom at site $j$. In this sense, $b_n(j)$ is a measure of local chirality.

As $b_n(j)$ transforms exactly the same way as a loop current circulating along the associated Cu-O-O triangle, we expect that an ansatz with $b_n(j) \neq 0$ has non-zero loop currents. Indeed, starting from a fully local theory and treating the hybdizations $t$ and $t'$ between neighboring atoms as a perturbation, we find (see \appref{SmallHybr}) as leading non-vanishing contribution 
\begin{subequations}
\begin{equation}
J^{\text{O-O}}_n(j) = t^2t'f_b\left(|\vec{H}_j|,\Delta,\mu\right) b_n(j) \label{JOO}
\end{equation}
for the current between two O atoms along the direction of $V_n$ in \figref{PhaseConvetions} and
\begin{equation}
J_n^{\text{O-Cu}}(j) = t^2t'f_b\left(|\vec{H}_j|,\Delta,\mu\right) \left(b_{n-1}(j)-b_{n}(j)\right) \label{JOCu}
\end{equation}\label{Currents}\end{subequations}
for the current from the O to the Cu atom along the bond associated with $W_n$. To write the expression for $J_n^{\text{O-Cu}}$ in form of a single equation, we have, again, made the cyclic identification $b_{0}\equiv b_{4}$. In \equref{Currents}, the dependence on the on-site energy scales $\vec{H}_j$, $\Delta$, and $\mu$ is described by the function $f_b(H,\Delta,\mu)$, given explicitly in \appref{SmallHybr}. As we are at this point only interested in establishing a direct relation between the quantities $\{b_n\}$ and the loop currents rather than using \equref{Currents} to calculate the currents quantitatively, we do not go beyond leading order in the hybridization $t$, $t'$.

The condition for having intra-unit-cell loop currents, \textit{i.e.}, no net current flow between different unit cells reads
\begin{equation}
J_n^{\text{O-Cu}}(j) + J_n^{\text{O-O}}(j) - J_{n-1}^{\text{O-O}}(j)=0. \label{BlochTheorem}
\end{equation}
Inserting the expressions for the current given in \equref{Currents}, we find that this condition is indeed satisfied for any value of $b_n(j)$. Note, however, this is a consequence of the perturbative treatement of hopping up to third order which only takes into account the intra-unit cell operators $b_n(j)$. As we are interested in the physics of loop currents in the presence of translational symmetry, we will focus on $b_n(j)=b_n$ in the following. In that case, Bloch's theorem \cite{PhysRev.75.502,doi:10.1143/JPSJ.65.3254} requires \equref{BlochTheorem} to hold exactly (to any order in $t$, $t'$).

Based on the relation between the orbital currents and the four independent quantities $b_n$ in \equref{Currents}, we will next classify the different translation-invariant, intra-unit-cell loop-current patterns of the three-orbital model. The discussion of possible ans\"atze for the chargon Hamiltonian (\ref{EffectiveChargonHam}) that realize these current patterns will be postponed to the \secref{PossibleAnsaetze} below.

To organize the presentation, let us first focus on configurations that break the two-fold rotation symmetry $C_2$ perpendicular to the plane but preserve the combined symmetry operation $\Theta C_2$ of two-fold rotation and time-reversal. This is the situation, we had focused on in our earlier work in the one-orbital model \cite{PhysRevLett.119.227002}.
As is readily seen from \tableref{TransformOfeb}, invariance under $\Theta C_2$ imposes the constraint $b_{n}=-b_{n+2}$. There are thus two independent $b_n$, say $b_1$ and $b_2$, which leads to three different cases to consider: First, $b_1 = \pm b_2$, which corresponds to current \textit{pattern A} shown in \tableref{LoopCurrentPattern} and is characterized by the additional residual symmetries $\sigma_{xz}$ (or $\sigma_{yz}$) and $\Theta\sigma_{yz}$ (or $\Theta\sigma_{xz}$ depending on the relative sign of $b_1$ and $b_2$). These symmetries impose the constraint $e_n=e_{n+1}$ (see \tableref{TransformOfeb}) on the time-reversal-invariant $e_n$, which will play an important role when discussing the behavior in the presence of a magnetic field in \secref{SymmetrySignatures}.
Assuming that the crystal structure preserves the reflection symmetry $\sigma_{xy}$ at the $xy$-plane, the associated three-dimensional magnetic point group is $m'mm$. The pattern has $2 \times 2 = 4$ domains, which are related by $C_4$ and correspond to the relative  sign of $b_1$, $b_2$ and to the two possible signs of $b_1$.

Second, the case $b_1 \neq 0$, $b_2 = 0$ (or, equivalently, $1\leftrightarrow 2$) corresponds to \textit{pattern B} in \tableref{LoopCurrentPattern}. It has the diagonal reflection symmetries $\sigma_{d'}$ and $\Theta \sigma_d$ (or $d \leftrightarrow d'$) and, again, four domains, corresponding to the global sign of $b_n$ and interchanging $b_1$ and $b_2$. In this case, symmetries only require $e_n=e_{n+2}$. 

In principle, there is also a third possible case defined by $b_1,b_2 \neq 0$ with $|b_1| \neq |b_2|$. However, it only preserves $\Theta C_2$ while all other in-plane symmetries are broken. In fact, the residual symmetry group is just the intersection of the symmetries of pattern A and B; the corresponding pattern can be regarded as a combination or mixture of these two patterns, which is why we do not add this case as an additional independent pattern in \tableref{LoopCurrentPattern}.

Let us proceed with the complementary case of patterns that preserve $C_2$ and are, hence, odd under $\Theta C_2$.
We first note that, while the previously discussed current patterns can be realized in the one-orbital model, see \tableref{LoopCurrentPattern}, this is not true for those that are even under $C_2$: on the square lattice, the current operator $J_{ij}$ associated with any bond $i\rightarrow j$ can be transformed into $J_{ji}=-J_{ij}$ by consecutive application of $C_2$ and an appropriately chosen translation operation and, hence, has to vanish. This is different in the three-orbital model, where translational symmetry does not act irreducibly but only within the set of Cu-$d$, O-$p_x$, and O-$p_y$ orbitals separately. This makes $C_2$-symmetric loop-current patterns possible in the three-orbital model as we show next.

Invariance under $C_2$ demands $b_n=b_{n+2}$ and, hence, there are again only two independent $b_n$, which we choose to be $b_1$ and $b_2$ as before. If $b_1=-b_2$, \textit{pattern C} will be realized, which only has two domains (related by $C_4$ or time-reversal) corresponding to the two possible choices of the global sign of $b_n$. This state is characterized by the magnetic rotation symmetry $\Theta C_4$ (leading to $e_n=e_{n+1}$) and reflection symmetries at the $xz$- and $yz$-planes.
As opposed to pattern A, $b_1=b_2$ does not correspond to a different domain of pattern C but to a different pattern which we refer to as \textit{pattern D}. This pattern is special as it is the only configuration that does not exhibit any in-plane reflection symmetry (without composing with $\Theta$). Furthermore, it is the only pattern where the sum of all $b_n$ is non-zero and, in that sense, possesses a net chirality. There are two domains, related by time-reversal (or reflection), which correspond to the global sign of $b_n$. 
Finally, it is left to discuss $|b_1| \neq |b_2|$. Only $C_2$, $\Theta \sigma_d$, and, thus also, $\Theta \sigma_{d'}$ remain symmetries -- the magnetic symmetry group is the intersection of those of pattern C and D, which is expected since $|b_1| \neq |b_2|$ can be seen as the simultaneous presence of pattern C and D. This is why this case is not discussed as an independent pattern in \tableref{LoopCurrentPattern} and in the analysis in the remainder of the paper. Note that $b_1\neq 0$, $b_2=0$ (or $1 \leftrightarrow 2$ for that matter) is not special from a symmetry point of view as it has the same magnetic point group as a generic configuration with $|b_1| \neq |b_2|$.

In principle, one can also consider the situation where both $C_2$ and $\Theta C_2$ are broken. However, these configurations can be viewed as combinations of the patterns A, B ($C_2$ odd) and C, D (odd under $\Theta C_2$) and will, hence, not be discussed further. This can be easily seen by noting that the combinations of $b_n$ corresponding to the four different loop-current patterns in \tableref{LoopCurrentPattern} represent four linearly independent basis vectors spanning the space of possible configurations of the four different quantities $b_n$.

\subsubsection{Possible ans\"atze}
\label{PossibleAnsaetze}
After classifying the different orbital-current patterns, as summarized in \tableref{LoopCurrentPattern}, and discussing their relation to the gauge-invariant quantities $b_n$ defined in \equref{DefinitionOfbs}, let us next analyze which configurations of or ans\"atze for $\vec{H}_j$, $U_{ij}$ give rise to the different current patterns. As one might expect, there are many, gauge-nonequivalent, ans\"atze with the same symmetries and current signatures which can be classified using the projective symmetry group approach \cite{Wen2}. 

To restrict the number of possible ans\"atze, we focus on states that are close to the fractionalized antiferromagnet, $U_{ij}=\mathds{1}$, $\vec{H}_j= H_0 (-1)^{j_x+j_y} \vec{e}_x$ (or any gauge-equivalent representation for that matter), which preserves all symmetries of the square lattice and time reversal. More precisely, we look for a family of ans\"atze that can be continuously deformed into the fractionalized antiferromagnet by tuning a set of parameters, denoted by $\epsilon_j$ in the following, to zero.
This is motivated by the proximity of long-range antiferromagnetism to the pseudogap state, by the good agreement of the spectrum of this ansatz with photoemission data, and the agreement of many properties of this ansatz with dynamical mean-field theory and quantum Monte Carlo data on the strongly coupled Hubbard model \cite{Scheurer201720580,PhysRevX.8.021048}.
For sufficiently small $\epsilon_j \ll 1$, these important consistency conditions are still satisfied and the finite values of $\epsilon_j$ induce the additional symmetry breaking and loop-current order. Note that, in the limit $\epsilon_j \ll 1$, the energy scale of the time-reversal-symmetry-breaking orbital currents is much smaller than $H_0$ and, hence, than the anti-nodal gap; this is consistent with numerical studies \cite{GreiterThomale1,GreiterThomale2,DevereauxLC} of finite clusters of the three-orbital model that yield upper bounds on orbital currents that are much smaller than the pseudogap. 

Our starting point for all four different patterns is the canted-N\'eel-like Higgs-field configuration,
\begin{subequations}
\begin{equation}
    \vec{H}_j = H_0 \left[ (-1)^{j_x+j_y} \vec{e}_x + \epsilon_1 \vec{e}_z \right], \label{CantedNeelAnsatz}
\end{equation}
where the small canting, $\epsilon_1 \ll 1$, has been introduced to conveniently discuss U(1), $\epsilon_1=0$, and $\mathbb{Z}_2$, $\epsilon_1\neq 0$, topological order simultaneously. 

For $U_{ij}=\mathds{1}$ on all bonds $i,j$, time-reversal and all space-group symmetries of the crystal are preserved, leading to $b_n(j)=0$. Let us, thus, instead consider the more general form
\begin{equation}
    W_n(j) = \mathds{1}, \quad V_n(j) = e^{-i s_n \epsilon_2  (-1)^{j_x+j_y} \sigma_x}, \label{OnlyUijPartOfAnsatz}
\end{equation}\label{AnsatzForThreeOrbModel}
\end{subequations}
with $s_n=\{+1,-1,0\}$, which yields
\begin{equation}
    e_n(j) = 2\cos(s_n \epsilon_2), \qquad b_n(j) = 2 H_0 \sin(s_n \epsilon_2),
\end{equation}
independent of $j$ as required by translational symmetry. 
Choosing $s_n$ equal to the sign of the non-zero values of $b_n$ listed in \tableref{LoopCurrentPattern} and $s_n=0$ for $n$ with $b_n=0$, the ansatz (\ref{AnsatzForThreeOrbModel}) will reproduce the correct symmetry signatures in $e_n(j)$ and $b_n(j)$ for all four different patterns A--D. One can further show that all indicated magnetic symmetries are preserved. Consider pattern D, where $V_n(j) = e^{-i \epsilon_2  (-1)^{j_x+j_y} \sigma_x}$, as an example. Possible gauge transformations accompanying the symmetry transformations of translation $T_\mu$ by $\vec{e}_\mu$, $\mu=x,y$, four-fold rotation $C_4$, and the magnetic reflection $\Theta \sigma_{xz}$, are $G_{T_\mu}=i\sigma_z$, $G_{C_4}=\mathds{1}$, and $G_{\Theta \sigma_{xz}}=i\sigma_z$, respectively.

This shows that \equref{AnsatzForThreeOrbModel} provides an ansatz that is continously connected to the fractionalized antiferromagnet, $(\epsilon_1,\epsilon_2)=(0,0)$, but restricts the symmetries to the magnetic space group of the different pattern A--D once $\epsilon_2$ is non-zero. A finite value of $\epsilon_1$ reduces the residual gauge symmetry from U(1) to $\mathbb{Z}_2$, gapping out the U(1) ``photon'' that is present for $\epsilon_1 = 0$.
We have also checked explicitly by diagonalization of the tight-binding Hamiltonian in \equref{EffectiveChargonHam} that these configurations reproduce the corresponding loop-current and kinetic-energy patterns depicted in \tableref{LoopCurrentPattern}.

Note that the class of ans\"atze in \equref{AnsatzForThreeOrbModel} has, in general, no associated conventional (on-site) magnetic order parameter: it is not possible to choose a gauge such that $U_{ij}=\mathds{1}$ on all bonds $i,j$ and all non-trivial aspects of the ansatz are contained in the Higgs-field texture. If this was possible, the condensation of the spinons with $\braket{R_{sj}} \propto \mathds{1}$ in this gauge would transform the effective chargon Hamiltonian (\ref{EffectiveChargonHam}) into that of electrons in the presence of long-range order with conventional magnetic order parameter $\braket{\Phi_i}$ following the texture of the Higgs field. This follows by noting that $\braket{R_{sj}} \propto \mathds{1}$ implies a trivial relation between chargons and electrons, see \equref{Trafo}, and between the Higgs field and $\braket{\Phi_i}$, according to \equref{HiggsOrderParamRelation}. Instead, the inevitable presence of non-trivial $U_{ij}$ in the gauge where $\braket{R_{sj}} \propto \mathds{1}$ leads to an electronic Hamiltonian with effective hoppings that are non-trivial in spin, \textit{i.e.}, involve some form of spin-orbit coupling. More specifically, the phase with $\braket{R_{sj}} \propto \mathds{1}$ of the three-orbital ansatz in \equref{AnsatzForThreeOrbModel} can be thought of as having (canted) N\'eel order on Cu atoms which does not break any symmetry in any spin-rotation invariant observable. The additional symmetry breaking results from the interplay with the oxygen atoms, which have no local moments, but non-trivial spin-spin correlations between neighboring atoms. We will come back to this interpretation in the context of spin models in \secref{SpinModels} below.       

In \appref{PossibleHiggsPhases}, we prove that there is no ansatz (even when including those that are not close to the antiferromagnet) for \equref{EffectiveChargonHam} with $U_{ij}=\mathds{1}$ and the symmetries of pattern C or D and, thus, no conventional on-site magnetic order parameter. This is a consequence of the restrictions arising from the preserved translational and rotation symmetries.
Note, however, that the loop current patterns A and B can be represented by $U_{ij}=\mathds{1}$ when $\vec{H}_{j}$ assumes the form of a conical spiral,
\begin{equation}
  \vec{H}_i = H_0 \left(\cos(\vec{Q}\vec{r}_i) ,\sin(\vec{Q}\vec{r}_i),\epsilon_1 \right)^T, \label{ConicalSpiral}
  \end{equation}  
with incommensurate $\vec{Q}$ and non-zero canting $\epsilon_1 > 0$. As has already been discussed in \refcite{PhysRevLett.119.227002} for the one-orbital model, pattern A and B correspond, respectively, to $\vec{Q} = (\pi-\epsilon_2,\pi)^T$ and $\vec{Q} = (\pi-\epsilon_2,\pi+\epsilon_2)^T$ with incommensurate $\epsilon_2$.

Similar to the ansatz (\ref{AnsatzForThreeOrbModel}) discussed above, the conical spiral also has two independent small parameters, $\epsilon_1$ and $\epsilon_2$, deforming the fractionalized antiferromagnet. However, for the conical spiral,  loop current order is inevitably tied to the reduction of the residual gauge group to $\mathbb{Z}_2$; time-reversal-symmetry will only be broken if $\epsilon_1$ and $\epsilon_2$ are both non-zero.
This is different for \equref{AnsatzForThreeOrbModel} which allows for loop currents with both U(1) and $\mathbb{Z}_2$ topological order.

%===================================================================================================================
\subsection{One-orbital model}
\label{OneOrbitalModel}
Let us now discuss the one-orbital model that only involves the Cu-$d$ orbitals forming a square lattice. In analogy to our discussion above, we consider an effective chargon Hamiltonian,
 \begin{align}\begin{split}
H_\psi =& - {\sum_{i,j}}^{\prime}\sum_{\alpha,\beta} t_{ij} \psi^\dagger_{i\alpha} \left(U_{ij}\right)_{\alpha\beta} \psi^\pdagger_{j\beta}  \\ & - {\sum_{j}}^{\prime}\sum_{\alpha,\beta}  \psi^\dagger_{j\alpha} \vec{\sigma}_{\alpha\beta} \psi^\pdagger_{j\beta} \cdot \vec{H}_{j}, \end{split} \label{EffectiveChargonHamOneOrbital}
\end{align}
and look for possible ans\"atze, $U_{ij}\in \text{SU}(2)$ and $\vec{H}_j \in\mathbb{R}$, that lead to the symmetries of the four different pattern A--D in \tableref{LoopCurrentPattern}. The sole difference compared to our analysis in \secref{ThreeOrbiModLatt} is that \equref{EffectiveChargonHamOneOrbital} now only involves the Cu sites on the square lattice as indicated by the prime in the sums over lattice sites. We assume that at least the nearest and next-to-nearest-neighbor hopping amplitudes are non-zero.

As analyzed in detail in \refcite{PhysRevLett.119.227002} and already mentioned above, the conical spiral Higgs texture in \equref{ConicalSpiral} along with $U_{ij}=\mathds{1}$ constitutes a possible ansatz for the loop-current patterns A and B in the one-orbital model.
In this section, we look for possible ans\"atze that can also realize the symmetries of pattern C and D.
Recall, however, that the preserved two-fold rotation symmetry $C_2$ of pattern C and D does not allow for non-zero loop currents in the one-orbital model. Nonetheless, one can ask the question whether it is possible to write down an ansatz for the square-lattice chargon Hamiltonian that leads to the symmetries of pattern C and D; the resulting theory can be seen as a minimal description of the corresponding loop-current phases. The different non-trivial magnetic point groups can still have physical consequences since there are, as we will show below, other observables in the Hilbert-space of the one-orbital model that probe the reduction of symmetry to the magnetic space groups of pattern C and D.

\subsubsection{Ans\"atze for pattern C and D}\label{AnsaetzeWithUs}
Based on the result of \appref{PossibleHiggsPhases} that there are no ans\"atze with $U_{ij}=\mathds{1}$ for pattern C and D and noting that any small deformation of $\vec{Q}$ in \equref{ConicalSpiral} from the antiferromagnetic value $\vec{Q}=(\pi,\pi)^T$ will necessarily break both $C_4$ and $\Theta C_4$,
we start again from the canted N\'eel configuration (\ref{CantedNeelAnsatz}) and look for small deformations of $U_{ij}=\mathds{1}$ that lead to the correct symmetry breaking. 

Motivated by the success of \equref{AnsatzForThreeOrbModel} for the three-orbital model, let us consider 
%\begin{equation}
 %   U_{j,j+\vec{\eta}^{(p)}_\mu} = (U_{j+\vec{\eta}^{(p)}_\mu,j})^\dagger = e^{- i g_\mu  \eta_2 (-1)^{j_x+j_y} \sigma_x}
%\end{equation}
\begin{align}
    U_{ij}=(U_{ji})^\dagger = \begin{cases} e^{- i s_\mu  \epsilon_2 (-1)^{j_x+j_y} \sigma_x}, \quad & j=i+\vec{\eta}^{(p)}_\mu,  \\ \mathds{1}, \quad & \text{otherwise}, \end{cases} \label{UijAnsatz}
\end{align}
with $\mu=1,\dots, N_p$ labeling the $N_p$ distinct $p$th-nearest-neighbor vectors denoted by $\vec{\eta}^{(p)}_\mu$, where we only include one of $\{\vec{\eta}^{(p)}_\mu,-\vec{\eta}^{(p)}_\mu\}$ to the list. For example, we have $N_1=2$ with $\vec{\eta}^{(1)}_1=\vec{e}_x$ and $\vec{\eta}^{(1)}_2=\vec{e}_y$. 
The additional parameters $s_\mu$ will be chosen so as to lead to the correct symmetries and $\epsilon_2\ll 1$ guarantees that $U_{ij}$ are close to $\mathds{1}$.

We first note that the ansatz in \equsref{CantedNeelAnsatz}{UijAnsatz} preserves translational symmetry [with $G_{T_\mu}=i\sigma_z$, $\mu=x,y$ in \equref{SpaceGroupTrafo}] for any choice of $s_\mu$ and $p$. It can also induce time-reversal-symmetry breaking, which is most easily seen for the case $\epsilon_1 \neq 0$ \footnote{For $\epsilon_1 = 0$, it still breaks time-reversal symmetry except for $p=1$, $s_1=s_2$. This case, however, does not play any role for our analysis as we focus on $s_1=-s_2$ for $p=1$.}: $\vec{H}_i \rightarrow - \vec{H}_i$ can only be ``undone'' by performing the gauge transformation $G_\Theta = \pm i\sigma_y$ which, however, leads to $\epsilon_2 \rightarrow -\epsilon_2$. 

Due to the alternating sign in the exponent of \equref{UijAnsatz}, $C_2$ is preserved if and only if $(\vec{\eta}_\mu^{(p)})_x+(\vec{\eta}_\mu^{(p)})_y$ is odd, i.e., if the $p$th nearest neighbor hopping connects different sublattices of the square lattice; this leads us to $p=1$ and $p=4$ as the two smallest values of $p$ satisfying this constraint.

To begin with $p=1$, note that $\sigma_{xz}$ is automatically preserved and, hence, pattern D cannot be described. However, the symmetries of pattern C are indeed realized upon choosing $s_1=-s_2$ for $\vec{\eta}^{(1)}_1=\vec{e}_x$ and $\vec{\eta}^{(2)}_1=\vec{e}_y$. To illustrate the symmetries of this phase, let us consider the Hermitian and gauge-invariant operators
\begin{subequations}
\begin{align}
    L^e_{s,s'}(j) &= \text{Tr}\left[U_{j,j+s\vec{e}_x} U_{j+s\vec{e}_x,j+s'\vec{e}_y} U_{j+s'\vec{e}_y,j}\right] \\
    L^b_{s,s'}(j) &= i H_j^a \,\text{Tr}\left[\sigma_a U_{j,j+s\vec{e}_x} U_{j+s\vec{e}_x,j+s'\vec{e}_y} U_{j+s'\vec{e}_y,j}\right],
\end{align}\label{NewLsOp}\end{subequations}
in analogy to $e_n(j)$ and $b_n(j)$ in \equref{LeadingWilsonLoops}.
Defining 
\begin{align}\begin{split}
 l^\mu_1(j) =  L^\mu_{+,+}(j), \qquad &  l^\mu_2(j) =  -L^\mu_{-,+}(j), \\
  l^\mu_3(j) =  L^\mu_{-,-}(j),\qquad & l^\mu_4(j) =  -L^\mu_{+,-}(j). \label{DefinitionOfNewBEs}
\end{split}\end{align}
for $\mu=e,b$, we find that $l_n^e$ and $l_n^b$ transform exactly as $e_n$ and $b_n$ (see \tableref{TransformOfeb}).
Inserting the ansatz in \equsref{CantedNeelAnsatz}{UijAnsatz}, we find 
\begin{align}\begin{split}
    &l^e_n(j) = 2 \cos(\epsilon_2(s_1-s_2)), \\
    &l^b_1(j) = l^b_3(j) = -l^b_2(j) \\ &\qquad = - l^b_4(j) = 2 H_0 \sin(\epsilon_2(s_1-s_2)),
\end{split}
\end{align}
which confirms our symmetry analysis. Note that, exactly as in case of the three-band model ansatz (\ref{AnsatzForThreeOrbModel}), the symmetries are independent of whether $\epsilon_1=0$ or $\epsilon_1 \neq 0$ -- the latter only determines whether the resulting state has U(1) or $\mathbb{Z}_2$ topological order.

As opposed to the $N_1=2$ nearest-neighbor bonds, there are $N_4=4$ fourth-nearest-neighbor bonds which transform non-trivially under $\sigma_{xz}$. This essential geometric property of the bonds allows to write down an ansatz for pattern D: Using the conventions, $\vec{\eta}^{_{(4)}}_1 = 2\vec{e}_x+\vec{e}_y$, $\vec{\eta}^{_{(4)}}_2 = \vec{e}_x+2\vec{e}_y$, $\vec{\eta}^{_{(4)}}_3 = -\vec{e}_x+2\vec{e}_y$, and $\vec{\eta}^{_{(4)}}_4 = -2\vec{e}_x+\vec{e}_y$, we find that $s_1=s_3=-s_2=-s_4$ leads to the correct symmetries. In this case, the (renormalized) fourth-nearest-neighbor hopping amplitudes in the effective chargon Hamiltonian (\ref{EffectiveChargonHamOneOrbital}) are required to be non-zero as well. Similar to our discussion of pattern C above, we can write down observables of the form of \equref{DefinitionOfNewBEs}, this time involving $U_{ij}$ with $i$ and $j$ being fourth-nearest neighbors, to probe the magnetic point symmetries of pattern D. Also in this case, the symmetry breaking only requires non-zero $\epsilon_2$ and can both be realized with U(1), $\epsilon_1=0$, or $\mathbb{Z}_2$, $\epsilon_1 \neq 0$, topological order.

\subsubsection{Bi-local Higgs field}
Above, we have presented ans\"atze for both the one-orbital and three-orbital chargon Hamiltonian in \equsref{EffectiveChargonHamOneOrbital}{EffectiveChargonHam} that realize the symmetries of the different loop-current patterns and are close to the antiferromagnet. 
While those for pattern A and B have associated conventional on-site magnetic order parameters with the same symmetries, this is not the case for pattern C and D; the ansatz for the latter necessarily involved non-trivial $U_{ij}$ (in any gauge). 
In this subsection, we show that the extension of the form of the effective chargon Hamiltonian (\ref{EffectiveChargonHamOneOrbital}) to also include bi-local Higgs fields, that lie on the bonds of the lattice, allows for ans\"atze for pattern C and D with $U_{ij}=\mathds{1}$. The associated conventional magnetic order parameters involve both on-site and inter-site spin moments or, put differently, non-trivial form factors.

To be more explicit, we generalize the local Higgs-chargon coupling $H_{\text{int}} = \sum'_j \psi^\dagger_j \vec{\sigma} \psi_j \cdot \vec{H}_j$ in the second line of \equref{EffectiveChargonHamOneOrbital} to
%\begin{equation}
%H_{\text{int}} ={\sum_j}^\prime \psi^\dagger_j \vec{\sigma} \psi^\pdagger_j \cdot \vec{H}_j + {\sum_{i \neq j}}^\prime\sum_{\alpha=0}^3 \psi^\dagger_i \sigma_\alpha \psi^\pdagger_j \cdot H_{\alpha,ij}, \label{nonlocalHiggsCoupling}
%\end{equation}
\begin{subequations}
\begin{equation}
H_{\text{int}} = {\sum_j}^\prime \psi^\dagger_j \vec{\sigma} \psi^\pdagger_j \cdot \vec{H}_j + \Delta H_{\text{int}}, 
\end{equation}
with additional bi-local Higgs-chargon coupling 
\begin{equation}
\Delta H_{\text{int}} = {\sum_{i \neq j}}^\prime\sum_{\alpha=0}^3 \psi^\dagger_i \sigma_\alpha \psi^\pdagger_j \cdot H_{\alpha,ij}, 
\end{equation}\label{nonlocalHiggsCoupling}\end{subequations}
where $H_{\alpha,ij}=H^*_{\alpha,ji}$ (due to Hermiticity) and $\sigma_\alpha = (\sigma_0;\sigma_x,\sigma_y,\sigma_z)$. The extension to include also the identity matrix for the bi-local Higgs field $H_{\alpha,ij}$ is required by gauge invariance: The existence of $H_{\alpha,ij}'$ with
\begin{equation}
\sum_\alpha U_{sgi}^\dagger \sigma_\alpha  U_{sgj}^\pdagger \cdot H_{\alpha,ij} = \sum_\alpha \sigma_\alpha \cdot H_{\alpha,ij}', 
\end{equation}
for generic $U_{sgi}, U_{sgj} \in \text{SU}(2)$, requires non-zero $H'_{\alpha=0,ij}$. We imagine the terms in \equref{nonlocalHiggsCoupling} to arise from rewriting the coupling of electrons $c_i$ to  Hubbard-Stratonovich fields, $\vec{\Phi}_{i}$ (on-site magnetism), and $\Phi_{\alpha,ij}$ (bond charge, $\alpha=0$, and/or spin-density, $\alpha=x,y,z$, waves),
\begin{equation}
H_{\text{int}} ={\sum_i}^\prime c^\dagger_i \vec{\sigma} c^\pdagger_i \cdot \vec{\Phi}_i + {\sum_{i \neq j}}^\prime \sum_{\alpha=0}^3 c^\dagger_i \sigma_\alpha c_j \cdot \Phi_{\alpha,ij}, \label{BondSpinDensityWave}
\end{equation}
by fractionalizing the electronic operator into spinons, $R_{si}$, and chargons, $\psi_i$, according to \equref{Trafo} and introducing the bi-local Higgs field $H_{\alpha,ij}$ via
\begin{equation} \sum_{\alpha=0}^3 R_{si}^\dagger \sigma_\alpha R_{sj}^\pdagger  \Phi_{\alpha,ij} = \sum_{\alpha=0}^3 \sigma_\alpha \cdot H_{\alpha,ij}. \label{IntroductionOfHiggsFields}
\end{equation}
%Two comments are in order. First, note that already a pure charge fluctuation on the bonds ($\Phi_{\alpha,ij}=\delta_{\alpha,0}\Phi_{0,ij}$) contributes to all four components of the Higgs field $H_{\alpha,ij}$ which is required due to gauge invariance. The only restriction arises when the fields $\Phi_{\alpha,ij}$ satisfy a reality constraint. For instance, consider $\Phi_{\alpha,ij}=\delta_{\alpha,0}\Phi_{0,ij} \in \mathbbm{R}$ which leads to $H_{0,ij}\in\mathbbm{R}$ and $H_{x,ij},H_{y,ij},H_{z,ij}\in i\mathbbm{R}$ as readily follows from \equref{IntroductionOfHiggsFields}.
We note that bond charge and spin density waves, \equref{BondSpinDensityWave}, have been considered before, \textit{e.g.}, in the cuprates  \cite{SSLR13,Fujita14,ATSS15} and, more recently, proposed to be relevant for the correlated insulating state in twisted bilayer graphene \cite{TBGNonlocal}.

Here we will focus on ans\"atze with $U_{ij} = \mathds{1}$ on all bonds and consider different configurations of the local and bi-local Higgs fields. In that case, the condensation of the spinons with $\braket{R_{si}} \propto \mathds{1}$ leads to electrons in the presence of long-range on-site spin, $\braket{\vec{\Phi}_i}$, and inter-site spin/charge order, $\braket{\Phi_{\alpha,ij}}$, with the same spatial form as $\vec{H}_i$ and $H_{\alpha,ij}$, see \equsref{HiggsOrderParamRelation}{IntroductionOfHiggsFields}, respectively.

As \equref{nonlocalHiggsCoupling} allows for many different ans\"atze, we organize our search by focusing on Higgs fields $H_{\alpha,ij}$ with small bond length $|i-j|$. As explained above, we are interested in states close to the fractionalized antiferromagnet, \textit{i.e.}, $U_{ij}=\mathds{1}$, $\vec{H}_i= H_0 (-1)^{i_x+i_y} \vec{e}_x$, and $H_{\alpha,ij}=0$ (or gauge-equivalent).
For the same reason as before, we choose the on-site Higgs field to have a canted N\'eel texture using the parameterization given in \equref{CantedNeelAnsatz}. Including the small canting, $\epsilon_1 \ll 1$, allows us to conveniently study U(1) and $\mathbb{Z}_2$ topological order at the same time.
Without further terms, this ansatz preserves all square-lattice symmetries and time-reversal. 
We next discuss the form of the bi-local Higgs terms $H_{\alpha,ij}$ with shortest bond length $|i-j|$ that have to be added to this ansatz in order to yield the symmetries of pattern C and D. 

For pattern C, adding Higgs terms on nearest-neighbor bonds already suffices. To see this, consider
\begin{equation}
\Delta H_{\text{int}} = i  \epsilon_2 H_0  {\sum_j}^{\prime} (-1)^{j_x+j_y} \sum_{\mu=x,y} s_\mu   \psi^\dagger_{j+\vec{e}_\mu} \sigma_x \psi^\pdagger_j  + \text{H.c.}. \label{Option1}
\end{equation}
with $s_\mu\in \mathbb{R}$.
Under $\sigma_{xz}$ and $C_4$, it holds $s_\mu \rightarrow s_\mu$ and $(s_x,s_y) \rightarrow (s_y,s_x)$, respectively. The magnetic symmetries of pattern C are, thus, realized when $s_x=-s_y$. The state has U(1) or $\mathbb{Z}_2$ topological order depending on whether $\epsilon_1=0$ or $\epsilon_1 \neq 0$. 

Note that, in linear order in $\epsilon_2$, this ansatz is mathematically equivalent to that chosen in \secref{AnsaetzeWithUs} for pattern C. However, the current consideration via orientational averaging of an inter-site magnetic moment provides a different physical picture for its microscopic origin.
Furthermore, there are also additional ans\"atze which are mathematically distinct from those possible in \equref{EffectiveChargonHamOneOrbital}: Consider the (third-nearest-neighbor) term
\begin{equation}
\Delta H_{\text{int}} =   \epsilon_2 H_0 {\sum_j}^{\prime} (-1)^{j_x+j_y} \sum_{\mu=x,y} s_\mu   \psi^\dagger_{j+2\vec{e}_\mu} \sigma_y \psi^\pdagger_j  + \text{H.c.}, \label{Option2}
\end{equation}
again with $s_\mu \in \mathbb{R}$, which leads to the symmetries of pattern C upon choosing $s_x=-s_y$. In this case, $\epsilon_1 \neq 0$ is required as otherwise time-reversal symmetry would not be broken.

Before proceeding with pattern D, let us illustrate the broken symmetries of these ans\"atze using observables constructed from $U_{ij}$ and the Higgs fields. To this end, consider
\begin{equation}
    \mathcal{L}_{\vec{\eta}}(j) := i \sum_{a=1}^4\sum_{\alpha=0}^3 \text{Tr} \left[\sigma_a \sigma_\alpha U^P_{j+\vec{\eta},j} \right] H_{a,j} H_{\alpha,j,j+\vec{\eta}} +\text{H.c.}, \label{Observable1}
\end{equation}
where $U^P_{j+\vec{\eta},j}$ represents a product of $U_{ij}\neq 0$ connecting the two square-lattice sites $j$, $j+\vec{\eta}$ and transforming as $U^P_{j+\vec{\eta},j} \rightarrow  U_{sg j+\vec{\eta}} U^P_{j+\vec{\eta},j}U_{sg j}^\dagger$ under the gauge transformations in \equref{SUGauge}. Being Hermitian and gauge invariant, $\mathcal{L}_{\vec{\eta}}(j)$ is an observable and easily seen to be odd under time reversal. For our ansatz in \equref{CantedNeelAnsatz} supplemented with the bi-local term (\ref{Option1}), we find $\mathcal{L}_{\vec{e}_\mu}(j)=\mathcal{L}_{-\vec{e}_\mu}(j)=4H^2_0 \epsilon_2 s_\mu$, independent of $j$ as required by translational symmetry. We further see that the magnetic point symmetries of pattern C are realized for $s_x=-s_y$ and that $\epsilon_1 \neq 0$ is not required, in accordance with our projective-symmetry discussion above.

To illustrate the broken symmetries resulting from supplementing \equref{CantedNeelAnsatz} by the bi-local term in \equref{Option2}, the observable $\mathcal{L}_{\vec{\eta}}(j)$ is not sufficient, which follows from the fact that it only involves one on-site Higgs field but the resulting symmetry-breaking term must vanish if either of the two terms in \equref{CantedNeelAnsatz} is zero as we have seen above. For this reason, we instead consider
\begin{align}\begin{split}
    \Delta_{\vec{\eta}_1,\vec{\eta}_2}(j) &:= i \sum_{a,b=1}^3 \sum_{\alpha=0}^3 \text{Tr} \left[ \sigma_a \sigma_\alpha U^P_{j+\vec{\eta}_1,j+\vec{\eta}_2}\sigma_b U^P_{j+\vec{\eta}_2,j} \right] \\
    & \qquad\qquad  \times H_{a,j} H_{\alpha,j,j+\vec{\eta}_1} H_{b,j+\vec{\eta}_2} + \text{H.c.}, \label{Observable2}
\end{split}\end{align}
which is, again, Hermitian, gauge invariant, and odd under time-reversal. This observable allows to probe the broken symmetries induced by \equref{Option2} upon properly choosing $\vec{\eta}_1$ and $\vec{\eta}_2$; we find $\Delta_{2\vec{e}_\mu,\vec{e}_\mu} = \Delta_{-2\vec{e}_\mu,-\vec{e}_\mu} = -8H^3_0\epsilon_1\epsilon_2 s_\mu$, which is only non-zero if $H_0,\epsilon_1,\epsilon_2,s_\mu \neq 0$ and transforms as the loop current pattern C under all magnetic symmetry operations (upon choosing $s_x=-s_y$ as discussed above). 

Pattern D can be analyzed in a similar way.
As expected based on our analysis in \secref{AnsaetzeWithUs}, the bi-local Higgs ansatz for pattern D that can, to leading order in $\epsilon_2$, be recast in the form of the ansatz in \equref{UijAnsatz}, is the fourth-nearest-neighbor term,
\begin{equation}
\Delta H_{\text{int}} = i \epsilon_2 H_0 {\sum_j}^{\prime} (-1)^{j_x+j_y} \sum_{\mu=1}^4 s_\mu   \psi^\dagger_{j+\vec{\eta}^{_{(4)}}_\mu} \sigma_x \psi^\pdagger_j  + \text{H.c.}. \label{Option3}
\end{equation}
%where $g_\mu \in \mathbbm{R}$. %and we defined the fourth-nearest-neighbor vectors  $\vec{\eta}^{_{(4)}}_1 = 2\vec{e}_x+\vec{e}_y$, $\vec{\eta}^{_{(4)}}_2 = \vec{e}_x+2\vec{e}_y$, $\vec{\eta}^{_{(4)}}_3 = -\vec{e}_x+2\vec{e}_y$, and $\vec{\eta}^{_{(4)}}_4 = -2\vec{e}_x+\vec{e}_y$. 
The symmetries of pattern D are correctly reproduced upon choosing $s_1=s_3=-s_2=-s_4$ with U(1) or $\mathbb{Z}_2$ topological order depending on whether we set $\epsilon_1=0$ or $\epsilon_1 \neq 0$.

Also in this case, we can write down a real bond order parameter, which is, hence, not of the same asymptotic mathematical form as those in \equref{UijAnsatz}.
However, it requires at least sixth-nearest-neighbor bonds,
\begin{equation}
\Delta H_{\text{int}} =  \epsilon_2H_0{\sum_j}^{\prime} (-1)^{j_x+j_y} \sum_{\mu=1}^4 s_\mu   \psi^\dagger_{j+\vec{\eta}^{_{(6)}}_\mu} \sigma_y \psi^\pdagger_j  + \text{H.c.}, \label{Option4}
\end{equation}
with sixth-nearest-neighbor vectors $\vec{\eta}^{_{(6)}}_1 = 3\vec{e}_x+\vec{e}_y$, $\vec{\eta}^{_{(6)}}_2 = \vec{e}_x+3\vec{e}_y$, $\vec{\eta}^{_{(6)}}_3 = -\vec{e}_x+3\vec{e}_y$, $\vec{\eta}^{_{(6)}}_4 = -3\vec{e}_x+\vec{e}_y$. Choosing again $s_1=s_3=-s_2=-s_4$ yields the symmetries of pattern D as long as $\epsilon_1\neq 0$.

Also in this case, we can probe the symmetries of these ans\"atze by evaluation of observables of the form of \equsref{Observable1}{Observable2}, once the appropriate bonds $\vec{\eta}$, $\vec{\eta}_{1,2}$ have been chosen: we find $\mathcal{L}_{\pm \vec{\eta}^{_{(4)}}_\mu}(j) = 4 \epsilon_2 s_\mu H_0^2$ and $\Delta_{\pm \vec{\eta}^{_{(6)}}_\mu,\pm \widetilde{\vec{\eta}}_\mu} = -8H_0^3\epsilon_1\epsilon_2 s_\mu$, where $\widetilde{\vec{\eta}}_1=-\widetilde{\vec{\eta}}_4=\vec{e}_x$, $\widetilde{\vec{\eta}}_2=-\widetilde{\vec{\eta}}_3=\vec{e}_y$, for the bi-local term in \equsref{Option3}{Option4}, respectively. We, thus, see explicitly that the symmetries of pattern D are correctly represented if $s_1=s_3=-s_2=-s_4$ as noted above.

%=================================================================================================================== 
\subsection{Spin degrees of freedom and \CP theory}
\label{CP1TheoryDescription}
So far, we have focused on the chargons, which was very natural given our motivation of describing topologically ordered states that exhibit non-zero orbital currents. In this section, we discuss the spin degrees of freedom. % which form a spin-liquid phase.
Based on our interest in states which are close to the antiferromagnet (see \secref{PossibleAnsaetze}), we will start from the \CP description of fluctuation antiferromagnetism \cite{CP1Action} and discuss various Higgs phases of the theory that lead to spin-liquid states with the same symmetries as the different loop-current patterns in \tableref{LoopCurrentPattern}.

\subsubsection{Spin models and loop currents}\label{SpinModels}
By design, any theory that only describes the spin degrees of freedom does not exhibit any currents. However, suitable combinations of spin operators can couple to the current operators in some order of $t/U$. As these operators have to be time-reversal odd and (in the absence of spin-orbit coupling) spin-rotation invariant, they must at least involve three spin operators. The only spin-rotation invariant combination of three spin operators $\hat{\vec{S}}$ is the triple product of three distinct spins (also known as scalar spin-chirality operators). Focusing on intra-unit-cell operators in the three-orbital model, we are left with 
\begin{equation}
\Delta_{s,s'}(i) := \hat{\vec{S}}_i \cdot \left( \hat{\vec{S}}_{i+s \frac{\vec{e}_x}{2}} \times \hat{\vec{S}}_{i+s' \frac{\vec{e}_y}{2}} \right), \quad s,s' = \pm,  \label{TripleProdcts}
\end{equation}
where, as before in \secref{ThreeOrbiModLatt}, integer (half-integer) indices refer to Cu (O) sites. In \refscite{CoupleToLoopCurrents}, it was explicitly demonstrated that spin-chirality operators can couple [at order $(t/U)^2$] to bond-current operators.

The scalar spin-chirality operators (\ref{TripleProdcts}) also resonate well with the interpretation of the ans\"atze in \equref{AnsatzForThreeOrbModel} for the loop-current phases in the three-orbital model discussed in \secref{PossibleAnsaetze}: The Higgs field (\ref{CantedNeelAnsatz}) on the Cu sites describes antiferromagnetism with non-zero magnitude of the local magnetic order but without long-range order, due to strong orientational fluctuations in the topologically ordered states (where $\braket{R_{si}}= 0$). While there is no Higgs field on the oxygen sites, $V_n$ in \equref{OnlyUijPartOfAnsatz} describes non-trivial local spin correlations $\hat{\vec{S}}_{i+s \frac{\vec{e}_x}{2}} \times \hat{\vec{S}}_{i+s' \frac{\vec{e}_y}{2}}$; while these vectors undergoe strong orientational fluctuations, too, leading to $\braket{\hat{\vec{S}}_{i+s \frac{\vec{e}_x}{2}} \times \hat{\vec{S}}_{i+s' \frac{\vec{e}_y}{2}}}=0$, its fluctuations are correlated with those on the Cu sites. This allows for non-zero expectation values of $\Delta_{ss'}$ in \equref{TripleProdcts}.

To connect more explicitly to the loop-operators $b_n$ in \equref{DefinitionOfbs}, that are related via \equref{Currents} to the current operators in the three-orbital model, let us define $\delta_n$, $n=1,2,3,4$, according to
\begin{align}\begin{split}
 \delta_1(i) =  \Delta_{+,+}(i), \qquad &  \delta_2(i) =  -\Delta_{-,+}(i), \\
  \delta_3(i) =  \Delta_{-,-}(i),\qquad & \delta_4(i) =  -\Delta_{+,-}(i). \label{DefinitionOfSs}
\end{split}\end{align}
With these definitions, $\delta_n(i)$ is found to have the same transformation properties as $b_n(i)$ under all symmetry operations (see \tableref{TransformOfeb}).
This not only provides a direct connection between the scalar spin-chirality operators (\ref{TripleProdcts}) and the Wilson loops $b_n$ of the gauge theory, but also allows to read off the spin operator that couples to the different loop-current patterns A--D from the symmetries of $b_n$ given in \tableref{LoopCurrentPattern}. 
For instance, in the case of pattern D, the associated spin-operator reads as
$\mathcal{O}^{S}_D = {\sum_{j}}^{\prime}\sum_{n=1}^4 \delta_n(j)$, which is $C_4$ symmetric, odd under time-reversal $\Theta$, and even under $\Theta \sigma_{xz}$. 

One can also write down observables with the same transformation properties as the four loop-current patterns A--D in terms of spin operators on the Cu atoms only, \textit{i.e.}, in the one-orbital model on the square lattice. This is achieved simply by replacing the oxygen atoms in \equref{TripleProdcts} by neighboring Cu sites,
 \begin{subequations}\begin{equation}
\widetilde{\Delta}_{s,s'}(i) := \hat{\vec{S}}_i \cdot \left( \hat{\vec{S}}_{i+s \vec{e}_x} \times \hat{\vec{S}}_{i+s' \vec{e}_y} \right), \quad s,s' = \pm,  \label{TripleProdcts2}
\end{equation}
and defining $\widetilde{\delta}_n$ analogous to \equref{DefinitionOfSs},  $\widetilde{\delta}_1(i) =  \widetilde{\Delta}_{+,+}(i)$ and so on. Also $\widetilde{\delta}_n$ transforms as $b_n$ under all symmetry operations and the order parameters for all different loop-current patterns follow from \tableref{LoopCurrentPattern} -- in particular, 
\begin{equation}
    \widetilde{\mathcal{O}}^{S}_D = {\sum_{j}}^{\prime}\sum_{n=1}^4 \widetilde{\delta}_n(j).
\end{equation}\label{OneOrbitalOP}\end{subequations}
The result of \appref{PossibleHiggsPhases} implies that there is no analogous classical magnetic texture on the square lattice, $\braket{\vec{S}_i}$, such that the classical factorization
\begin{equation}
    \braket{\widetilde{\Delta}_{s,s'}(i)}_{\text{cl}} = \braket{\hat{\vec{S}}_i} \cdot \left( \braket{\hat{\vec{S}}_{i+s \vec{e}_x}} \times \braket{\hat{\vec{S}}_{i+s' \vec{e}_y}} \right)
\end{equation}
of the expectation value of \equref{TripleProdcts2}
captures the symmetries of pattern C and D. In contrast, for pattern A and B, this is possible once $\braket{\vec{S}_i}$ assumes the form of a conical spiral \cite{PhysRevLett.119.227002}.

This crucial difference between pattern A/B and C/D will also be reflected in the \CP formalism: in \secref{PatternCAndD} below, we will find that the \CP description of phases C and D requires more complicated terms (with more fields and derivatives) than those of phases A and B. This will again be traced back to the presence of the magnetic point symmetry $C_2\Theta$ of the latter two loop current patterns while the loop currents C and D are odd under $C_2\Theta$. 
For completeness, we finally point out that this difference in symmetry has the following additional consequence for the description in terms of spin operators. Pattern A and B can couple to operators of the form
\begin{equation}
L_{\vec{\eta}}(i) =  \hat{\vec{S}}_i \cdot \left( \hat{\vec{S}}_{i+ \vec{\eta}} \times \hat{\vec{S}}_{i- \vec{\eta}} \right) ,
\end{equation}
with $\vec{\eta} =\vec{e}_x$ for pattern A and $\vec{\eta}=(1,-1)$ for pattern B. For any loop current pattern, such as C and D, with preserved $C_2$ (and translational) symmetry, we must have $\braket{L_{\vec{\eta}}}=0$ for any $\vec{\eta}$.

\newcommand{\extrasp}{\phantom{me'}}

\begin{table*}[bt]
\begin{center}
\caption{Representation of all relevant symmetries, translation $T_{x,y}$ by lattice spacing along $x,y$, spin-rotation SU(2)$_s$, time-reversal $\Theta$, two-fold rotation $C_2$, four-fold rotation $C_4$, and reflection $\sigma_{yz}$, in the \CP theory (\ref{StandardCP1Action}) with additional Higgs fields $P$, $Q_a$, and $W_{ab}$, defined in \equsref{LH}{LW}, respectively. We use $R_g$ to denote the representation of $g=\sigma_{yz},C_4$ on 2D coordinates $\vec{x}=(x,y)^T$. As $C_2$ plays a crucial role in the discussion of \secref{PatternCAndD}, we have added $C_2$ as a separate symmetry although $(C_4)^2 = C_2$.}
\label{SymmetriesCP1}
 \begin{tabular}{ccccc} \hline \hline
Symmetry & \CP fields &  $P$ & $Q_a$ & $W_{ab}$ \\ \hline
$T_{x,y}$ & $ z_\alpha(\vec{x},t) \rightarrow \varepsilon_{\alpha\beta}z^*_{\beta}(\vec{x},t)$  & $P \rightarrow P^*$ & $Q_a \rightarrow Q^*_a$ & $W_{ab} \rightarrow W^*_{ab}$  \\
\extrasp SU(2)$_s$ \extrasp & \extrasp $ z_\alpha(\vec{x},t) \rightarrow \left(e^{i \vec{\varphi}\cdot \vec{\sigma}}\right)_{\alpha\beta}z_{\beta}(\vec{x},t)$ \extrasp & \extrasp $P \rightarrow P$ \extrasp & \extrasp $Q_a \rightarrow Q_a$ \extrasp & \extrasp $W_{ab} \rightarrow W_{ab}$ \extrasp \\
$C_2$ & $ z_\alpha(\vec{x},t) \rightarrow z_{\alpha}(-\vec{x},t)$  & $P \rightarrow P$ & $Q_a \rightarrow -Q_a$ & $W_{ab} \rightarrow W_{ab}$ \\
$\Theta$ & $ z_\alpha(\vec{x},t) \rightarrow \varepsilon_{\alpha\beta}z_{\beta}(\vec{x},-t)$ &  $P \rightarrow -P$ & $Q_a \rightarrow Q_a$ & $W_{ab} \rightarrow W_{ab}$ \\ 
$C_4$ & $ z_\alpha(\vec{x},t) \rightarrow z_{\alpha}(R_{C_4}\vec{x},t)$  & $P \rightarrow P$ & $Q_a \rightarrow (R_{C_4}Q)_a$ & $W_{ab} \rightarrow (R_{C_4}W R^T_{C_4})_{ab}$ \\ 
$\sigma_{yz}$ & $ z_\alpha(\vec{x},t) \rightarrow z_{\alpha}(R_{\sigma_{yz}}\vec{x},t)$   & $P \rightarrow P$ & \extrasp $Q_a \rightarrow (R_{\sigma_{yz}}Q)_a$ \extrasp & \extrasp $W_{ab} \rightarrow (R_{\sigma_{yz}}WR^T_{\sigma_{yz}})_{ab}$ \extrasp \\ \hline \hline
 \end{tabular}
\end{center}
\end{table*}

\subsubsection{Pattern A and B}
Let us now turn to the \CP description of these phases and begin with pattern A and B. As has already been discussed in \refcite{PhysRevLett.119.227002}, the \CP theory naturally leads to phases with $\mathbb{Z}_2$ topological order and exactly the same symmetries as the loop-current pattern A and B. In this subsection, we briefly review and introduce the notation in order to describe the states with the symmetries of pattern C and D in \secref{PatternCAndD} below.

The \CP action of fluctuating antiferromagnetism reads as  \cite{CP1Action}
\begin{equation}
\mathcal{S} = \frac{1}{g} \int \diff^2 x \, \diff t \, | D_\mu z_\alpha|^2 + \dots, \,\, D_\mu = \partial_\mu - i a_\mu, \label{StandardCP1Action}
\end{equation}
where the integration and the index $\mu$ of the derivative $\partial_\mu$ involve two-dimensional space, $\vec{x}=(x,y)$, and time, $t$, $z_\alpha$ are two-component bosonic \CP fields (with constraint $z_\alpha^* z^{\phantom{*}}_\alpha=1$ and related to the local N\'eel order $\vec{n}$ according to $\vec{n}=z^\dagger \vec{\sigma} z$), and $a_\mu$ are emergent U(1) gauge fields. By virtue of being compact, the gauge fields allow for monopoles which require additional regularizations and Berry-phase terms represented by the ellipsis. However, these terms are not crucial for the following analysis as we focus on $\mathbb{Z}_2$ topologically ordered states where monopoles are suppressed. 

While we will mainly focus on symmetry arguments in this section, the \CP action can, in principle, also be derived \cite{Scheurer201720580,PhysRevLett.119.227002} from the microscopic Hubbard-like model by rewriting the electronic operators according to \equref{RPsi} and integrating out the chargons (technically only possible in the insulator); after rewriting the spinon fields $R_{si}$ as in \equref{Rz}, a gradient expansion yields the continuum \CP theory. 

The prefactor $1/g$ in \equref{StandardCP1Action} controls the strength of fluctuations; for small $g$, we obtain a conventionally ordered N\'eel phase, where $\braket{z_\alpha}, \braket{\vec{n}}\neq 0$, while large $g$ induces a gap to the bosons. Without further terms in the action, confinement will eventually lead to valence bond solid (VBS) order \cite{NRSS89,NRSS90}.

To avoid confinement for large $g$, we add charge-$2$ Higgs fields ($\sim z z$). As we are only interested in spin-rotation invariant Higgs phases (see transformation behavior of the \CP fields summarized in \tableref{SymmetriesCP1}) and $z_\alpha \varepsilon_{\alpha\beta} z_{\beta}=0$, with $\varepsilon_{\alpha\beta} = (i\sigma_y)_{\alpha\beta}$, the leading non-vanishing terms we can consider are $z_\alpha \varepsilon_{\alpha\beta}\partial_t z_{\beta}$ and $z_\alpha \varepsilon_{\alpha\beta}\partial_a z_{\beta}$, $a=x,y$. This motivates considering the extended action $\mathcal{S} \rightarrow \mathcal{S} + \int \diff^2 x \, \diff t \,\mathcal{L}_{P,Q}$ with
\begin{align}\begin{split}
\mathcal{L}_{P,Q} &=   |(\partial_\mu - 2 i a_\mu) P|^2 +  |(\partial_\mu - 2 i a_\mu) Q_a |^2  \label{LH}
 \\ &+ \left(\lambda_1 P^\ast \, \varepsilon_{\alpha\beta} z_\alpha
\partial_t z_\beta +  \lambda_2 Q_a^\ast  \varepsilon_{\alpha\beta} z_\alpha
\partial_a z_\beta  + \text{H.c.} \right)\\ &- m^2_P |P|^2 - m^2_Q |Q_a|^2  
 + \ldots  ,
\end{split}\end{align}
where the ellipsis stands for higher order terms in the Higgs potentials. From the transformation properties summarized in \tableref{SymmetriesCP1}, we can see that if both $P$ and $Q_a$ condense, time-reversal and two-fold rotation symmetry $C_2$ are broken while their product is preserved. Translational symmetry is present as long as $\braket{P}\braket{Q_a}^* \in \mathbb{R}$. This shows that phases with the symmetries of the loop current patterns A and B are obtained as Higgs phases of the quadratic \CP theory in \equsref{StandardCP1Action}{LH}: The symmetries of pattern A are obtained when
\begin{equation}
\braket{Q_x} = q, \quad\braket{Q_y} = 0, \quad \braket{P} = p,
\end{equation}
and those of pattern B if
\begin{equation}
\braket{Q_x} = -\braket{Q_y} = q, \quad \braket{P} = p,
\end{equation}
where $pq^*\in\mathbb{R}$.
An observable, \textit{i.e.}, a gauge-invariant, Hermitian operator, in terms of the Higgs fields that can couple to the current patterns A and B is given by
\begin{equation}
\mathcal{O}^{\CPm}_a = Q_a^* P + Q_a P^*.
\end{equation}
We have $\braket{\vec{\mathcal{O}}^{\CPm}}=2pq^*(1,0)$ and $\braket{\vec{\mathcal{O}}^{\CPm}}=2pq^*(1,-1)$ for pattern A and B, respectively.
In \refcite{PhysRevLett.119.227002}, these Higgs phases have been explicitly derived from the one-orbital SU(2) gauge theory of \secref{OneOrbitalModel}.

While the \CP theory very naturally leads to phases with the symmetries of pattern A and B, obtaining those of the current patterns C and D is more difficult, as discussed next. 

\subsubsection{Pattern C and D}
\label{PatternCAndD}
The \CP theory in \equsref{StandardCP1Action}{LH} cannot have the same symmetries as those of pattern C and D since it necessarily preserves the product of $C_2$ and time-reversal: As $Q_a$ and $P$ are even (odd) and odd (even) under time-reversal ($C_2$), we can only either preserve time-reversal and $C_2$ (only one of the fields condenses) or break both at the same time (both condense).  In fact, any quadratic \CP theory with spin-rotation and translational symmetry has the property that if time-reversal is broken, the same holds for $C_2$. To see this, first note that spin-rotation invariance only allows for two different types of terms,
\begin{equation}
 c_1\, z_\alpha^* \partial^n z^\pdagger_\alpha + \text{H.c.} \quad \text{and} \quad  c_2\, \varepsilon_{\alpha\beta} z_\alpha  \partial^n z_\beta + \text{H.c.}, \label{TwoTypesOfTerms}
 \end{equation} 
with, in general, complex prefactors $c_{1,2}\in \mathbb{C}$. Here $\partial^n$ represents $n$ derivatives with respect to either space or time or any mixture of the two.

To begin with the first term, translational symmetry requires $c_1 \in \mathbb{R}$. Consequently, only even powers $n$ can contribute (in the bulk) and, hence, this terms has to be invariant under time-reversal in order to preserve $C_2$ (and vice versa).   

In the second term in \equref{TwoTypesOfTerms}, only odd $n$ can contribute due to the antisymmetry of $\varepsilon$. As additional multiplication of $\partial^n$ by $\partial^2_a$, $\partial^2_t$, or $\partial_a\partial_t$ does not yield terms with different behavior under $\Theta$ or $C_2$, we can focus on $\partial^n=\partial_t$ or $\partial^n=\partial_a$, which are just the terms generated by condensation of the Higgs fields $P$ and $Q_a$ in \equref{LH}. This proves that $\Theta C_2$  is a symmetry of any local, quadratic \CP theory with spin-rotation and translational symmetry.

Consequently, we necessarily have to go beyond quadratic order to describe phases with the same symmetries as pattern C and D. As terms with three $z$ fields necessarily lead to the loss of topological order, we have to study expressions involving four $z$ fields (order parameters expressed in terms of gauge fields will be discussed at the end of this section). Naturally, there are many terms involving four bosonic fields that can be considered. Therefore, we first focus on charge-$2$ terms of the form $\sim z^*z^3$. Naively, there are two possible spin-rotation invariant Higgs candidates to consider, which, to leading order in derivatives, have the from
\begin{subequations}\begin{align}
  (z^*_\alpha \partial_{\mu} z^\pdagger_\alpha) \, (\varepsilon_{\beta\gamma} z_\beta^\pdagger \partial_{\mu'} z^\pdagger_{\gamma}), \label{FourZOption1} \\
  (z^*_\alpha  \vec{\sigma}_{\alpha\alpha'} \partial^n z^\pdagger_{\alpha'}) \cdot (\varepsilon_{\beta\gamma}\vec{\sigma}_{\gamma\delta} z_\beta^\pdagger \partial^{n'} z^\pdagger_{\delta}), \label{FourZOption2}
\end{align}\label{FourBosonTermsToConsider}\end{subequations}
where $\mu,\mu'=x,y,t$ and $n+n'=2$. 
However, it is easily seen that the terms of the form of \equref{FourZOption2} with $n=2$, $n'=0$ (and $n \leftrightarrow n'$) vanish (upon integrating by parts) and that the remaining possible case, $n=n'=1$, is equivalent to \equref{FourZOption1} which can be shown by partial integration. We can, hence, focus on \equref{FourZOption1}.
To further restrict the number of possible choices of $\mu$ and $\mu'$, we note that the total number of spatial derivatives of any Higgs term in phase C or D cannot be one. This results from the combination of translational and $C_4$ rotation (or $\Theta C_4$) symmetry: Take, \textit{e.g.}, $Q_a$ which we have studied earlier. Due to $C_4$ (or $\Theta C_4$) rotation symmetry, we have $\braket{Q_x} = \pm i \braket{Q_y}$ which, at the same time, is inconsistent with translational symmetry. The same argument applies to the four-boson terms in \equref{FourBosonTermsToConsider}. 

In combination with the fact that both current patterns C and D break mirror reflection symmetries, the minimal number of spatial derivatives is two. 
We are thus left with just a single term, with both derivatives in \equref{FourZOption1} being spatial, and, hence, extend the action in \equref{StandardCP1Action} according to $\mathcal{S} \rightarrow \mathcal{S} + \int \diff^2 x \, \diff t \,(\mathcal{L}_{P,Q}+\mathcal{L}_{W})$, where 
%\begin{align}\begin{split}
%\mathcal{L}_{W} &=   |(\partial_\mu - 2 i a_\mu) W_{ab}|^2 + V(W_{ab}) \\ &+ W^*_{ab} \, (z^*_\alpha \partial_a z^\pdagger_\alpha) \, (\varepsilon_{\beta\gamma} z_\beta^\pdagger \partial_b z^\pdagger_{\gamma}) + \text{H.c.},    \label{LW}
%\end{split}\end{align}
\begin{align}\begin{split}
\mathcal{L}_{W} &=   \left((\partial_\mu + 2 i a_\mu) W^*_{ab}\right) (\partial_\mu - 2 i a_\mu) W_{ba} + V(W_{ab}) \\ &+ \lambda_3 W^*_{ab} \, (z^*_\alpha D_a z^\pdagger_\alpha) \, (\varepsilon_{\beta\gamma} z_\beta^\pdagger \partial_b z^\pdagger_{\gamma}) + \text{H.c.},    \label{LW}
\end{split}\end{align} 
with $V(W_{ab})$ describing the Higgs potential and $a,b=x,y$ being summed over. 
Note that $W_{ab}$ can be restricted to be symmetric under $a\leftrightarrow b$ as the antisymmetric part of $W_{ab}$ only couples to a (quadratic) boundary term.

As follows from the symmetry representations listed in \tableref{SymmetriesCP1}, pattern C is obtained as
\begin{align}\begin{split}
\braket{Q_a} = 0, & \quad \braket{P} = p,  \\ \braket{W_{xx}}=-\braket{W_{yy}}=w, & \quad \braket{W_{xy}}=\braket{W_{yx}}=0,
\end{split}\end{align}
where, due to translational symmetry, $pw^* \in \mathbb{R}$.

As before, we can define observables in terms of Higgs fields that can directly couple to the loop currents. 
To this end, let $\mathcal{R}_{ab} = W^*_{ab}P + W_{ab}P^*$,
%\begin{equation}
%\mathcal{R}_{ab} = W^*_{ab}P + W_{ab}P^*
%\end{equation}
which is Hermitian, gauge invariant, spin-rotation symmetric, invariant under translation, and odd under time-reversal (see \tableref{SymmetriesCP1}). The combination 
\beq
\mathcal{O}^{\CPm}_C = \mathcal{R}_{xx} - \mathcal{R}_{yy} 
\eeq
of the different components of $\mathcal{R}_{ab}$ has exactly the same symmetries as the loop-current patterns C and can therefore be seen as the corresponding \CP order parameter. 

In the case of pattern D, however, rotation and reflections symmetries require $\braket{W_{xy}}=-\braket{W_{yx}}\neq 0$ (or, equivalently, $\mathcal{O}^{\CPm}_D = \mathcal{R}_{xy} - \mathcal{R}_{yx}$) which cannot be realized due to $W_{ab}=W_{ba}$ as discussed above.

So far, we have been focusing on charge-$2$ quartic Higgs fields, $\sim z^* z^3$. However, in principle, also U(1) symmetric, $\sim (z^*z)^2$, or charge-$4$ Higgs fields, $\sim z^4$, are conceivable. Interestingly, as is shown in  \appref{FurtherHiggsTerms}, these two additional classes of terms do not allow for the symmetries of pattern C and D with two or fewer derivatives. 

Consequently, in order to realize pattern D, we have to consider Higgs fields involving higher-order derivatives.
It can be realized, for instance, by extending \equref{LW} to include a charge-$2$ Higgs field $X_{ab}^\mu$ with coupling
\begin{equation}
    \lambda_4 \left(X^\mu_{ab}\right)^* \, (z^*_\alpha D_a D_\mu^2 z^\pdagger_\alpha) \, (\varepsilon_{\beta\gamma} z_\beta^\pdagger \partial_b z^\pdagger_{\gamma}) + \text{H.c.}, \label{CouplingToX}
\end{equation}
where, as before, $a,b=x,y$ and $\mu=x,y,t$. This allows to write down a \CP order parameter for pattern D, 
\beq
\mathcal{O}^{\CPm}_D = \sum_\mu P^*(X^\mu_{xy}-X^\mu_{yx}) + \text{H.c.}\,; \label{OCP1D}
\eeq
the symmetries of pattern D are realized, \textit{e.g.}, when $\braket{P}=p$ and $\braket{X_{xy}^x}=-\braket{X_{yx}^y} = x$ with $px^* \in \mathbb{R}$. 

As discussed in \appref{NonLinearSigmaModel}, very similar behavior is found in the semi-classical O(3) non-linear sigma model \cite{ssbook} description of pattern D: as a consequence of rotation and reflections symmetries, the O(3) non-linear sigma model expression for the order parameter $\widetilde{\mathcal{O}}^{S}_D$, defined in \equref{OneOrbitalOP}, has its first non-zero contribution at fourth order in the gradient expansion and its form closely parallels that of $\mathcal{O}^{\CPm}_D$. In contrast, the order parameter of pattern C can be represented involving two spatial gradients.

Finally, we mention that one can also write down an order parameter for pattern D in terms of the gauge fields $a_\mu$. 
As pointed out in \refcite{Wang18}, the operator $\varepsilon_{\mu\nu\lambda} \partial_\rho f_{\rho\mu}f_{\nu\lambda}$, with $f_{\mu\nu}=\partial_{\mu} a_\nu -\partial_\nu a_\mu$ and three-dimensional Levi-Civita symbol $\varepsilon_{\mu\nu\lambda}$, has the same transformation properties as $ \widetilde{\mathcal{O}}^{S}_D$ under all symmetries of the square lattice and $\Theta$; upon noting that $a_\mu$ transforms as $i z^\dagger \partial_\mu z$ [see \equref{StandardCP1Action}], it is readily seen that this is the gauge-invariant combination of gauge fields with the fewest number of fields and derivatives that has the same symmetries as pattern D. Also its structure is very similar to $\mathcal{O}^{\CPm}_D$ in Eq.~(\ref{OCP1D}) (with $a_\mu \sim i z^\dagger \partial_\mu z$, both operators contain five derivatives).    
Transforming the same way under all symmetries of the system [including the emergent continuous spatial rotation symmetry of the continuum \CP action], the gauge-field operator $\varepsilon_{\mu\nu\lambda} \partial_\rho f_{\rho\mu}f_{\nu\lambda}$ and $\mathcal{O}^{\CPm}_D$ will, in general, be coupled in the effective action obtained by integrating out the \CP bosons $z_\alpha$.

%=================================================================================================================== 
\section{Fermionic spinons and bosonic chargons}
\label{FermionicSpinons}

In this section, we turn to the fermionic spinon approach outlined in Section~\ref{sec:fermions}. Among the orbital current patterns illustrated in Table~\ref{LoopCurrentPattern}, the fermionic spinon realizations of patterns A and B were already presented in Ref.~\onlinecite{ATSS18}.
Motivated by the observation that the description of pattern D was most complicated in the approach with bosonic spinons, \textit{i.e.}, involved the largest number of derivatives or furthest neighbor hopping, we will focus on the case of pattern D in the following. %Here we will limit ourselves to case of pattern D, and will not consider pattern C.

One of the main results of our paper is that pattern D is connected to a spin liquid that has been extensively studied in the literature: 
the chiral spin liquid \cite{KL87,WWZ89}. We will recall the fermionic spinon theory of the chiral spin liquid in Section~\ref{sec:piflux} using the one-band square lattice model. However, the one-band formulation is not sufficient to detect the nature of the orbital currents: the four-fold rotational symmetry of the state implies that orbital currents vanish identically on all links of the one-band square lattice. We then proceed to the discussion of the three-band case in Section~\ref{sec:piflux3}, and demonstrate the presence of orbital currents in the pattern D.

\subsection{$\pi$-flux and chiral spin liquid states in the one-band model}
\label{sec:piflux}

For our purposes, it is convenient to set up the fermionic spinon formulation by starting with an underlying Hubbard model (rather than the more commonly
used $t$-$J$ model). We begin by writing the parameterization in Eq.~(\ref{FR}) in the form
\beq
c_{i \alpha} = b_{i1} f_{i \alpha}  + b_{i2}^\ast \, \varepsilon_{\alpha\beta} f_{i \beta}^\dagger
\label{cbf}
\eeq
where $\alpha = \uparrow, \downarrow$ is a spin index. Inserting this into the hopping terms of the Hubbard model, we use
\beq
c_{i \alpha}^\dagger c_{j \alpha} = \left( f_{i \alpha}^\dagger f_{j \alpha} \right) \left( b_{i1}^\ast b_{j1} \right) +
\left( f_{i \alpha} f_{j \alpha}^\dagger  \right) \left( b_{j2}^\ast b_{i2} \right) + \ldots
\label{fbhopping}
\eeq
The terms omitted in Eq.~(\ref{fbhopping}) involve $f$-fermion pair operators whose average values vanish in the states we consider below. For the local number density, after factorizing fermion and boson expectation values, Eq.~(\ref{fbhopping}) implies
\beq
\left\langle c_{i \alpha}^\dagger c_{i \alpha} \right\rangle = 1 + \left( \left\langle f_{i \alpha}^\dagger f_{i \alpha}\right\rangle -1 \right) \left\langle |b_{i1}|^2 - |b_{i2}|^2 \right\rangle \,.
\label{fbnumber}
\eeq

\begin{figure}
\centering
\includegraphics[width=3in]{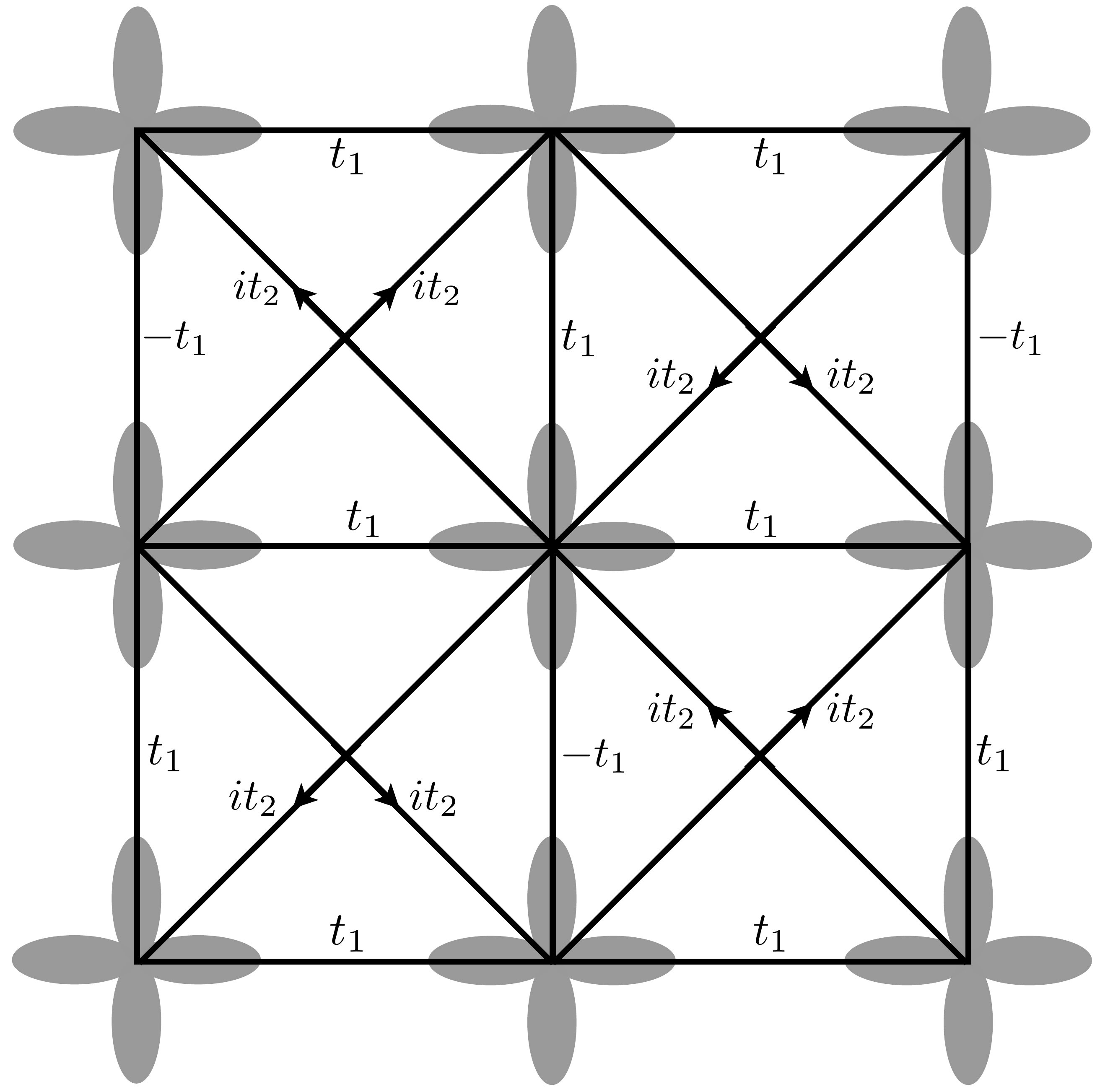}
\caption{Sketch of the saddle-point Hamiltonian, $H_f$, for the fermionic spinons, $f$, in the one-band model. The hopping parameters $t_1$ and $t_2$ are real. 
The $\pi$-flux state is obtained by $t_2=0$, and the chiral spin liquid for $t_2 \neq 0$.}
\label{fig:fermion}
\end{figure}

We now replace the boson operators in Eq.~(\ref{fbhopping}) by expectation values, and assume that they lead to an effective Hamiltonian
for the fermionic spinons, $f_\alpha$, of the form
\beq
H_f = - \sum_{i<j} \left( \bar{t}_{ij}  f_{i\alpha}^\dagger f_{j\alpha} + \bar{t}^\ast_{ij} f_{j\alpha}^\dagger f_{i\alpha} \right)\,.
\eeq
Note that the $\bar{t}_{ij}$ have been renormalized from the bare $t_{ij}$ in the electronic three-band model by factors of $\langle b_{i1}^\ast b_{j1} - b_{i2}^\ast b_{j2} \rangle$; the $t_{ij}$ are gauge-invariant, while the $\bar{t}_{ij}$ are not.
To obtain the $\pi$ flux and chiral spin liquid states, we take first and second neighbor hopping $\pm t_1 $ and $\pm i t_2$ as shown in Fig.~\ref{fig:fermion}.
We employ a 2 site unit cell, and then the momentum space Hamiltonian is
\beq
H_f = \sum_k \sum_{a,b} f_{a \alpha}^\dagger (k) M_{ab} (k) f_{b \alpha} (k) \label{HfM}
\eeq
where $a,b = A,B$ are sublattice indices, and the matrix $M$ is specified by
\bea
M_{11} &=& - M_{22} =  -2 t_2 \sin(k_x + k_y) -2 t_2 \sin(k_x - k_y) \nonumber \\
M_{12} &=& M_{21}^\ast = -2t_1 \cos(k_x) -2 i t_1 \sin (k_y)
\eea
This Hamiltonian has Dirac nodes at valleys $v=1$ at $\vec{k} = (\pi/2, 0)$ and $v=2$ at $\vec{k} = (-\pi/2, 0)$. 
We focus on the vicinities of these points by writing $\vec{k} = (\pm \pi/2 + q_x, q_y)$ and expand for small $q_x, q_y$. We also introduce Pauli matrices
$\tau^\lambda$ in sublattice space, $\sigma^\lambda$ in spin space, and $\mu^\lambda$ in valley space.  Then we can write the Hamiltonian as
\bea
M &= &  2 t_1 \mu^z \tau^x q_x  + 2 t_1  \tau^y  q_y - 4 t_2 \tau^z \mu^z
\eea
This is the Hamiltonian of two species of two-component Dirac fermions with mass $ 4 t_2$. 

Importantly, the chirality of the masses at the 2 Dirac nodes is the same. To see this, let us introduce the Lagrangian density associated with $M$
\beq
\mathcal{L} = f^\dagger \left[ \partial_\tau  - 2i t_1 \mu^z \tau^x \partial_x - 2 i t_1 \tau^y \partial_y  - 4 t_2 \tau^z \mu^z \right] f
\eeq
We want to remove the $\mu^z$ in the $x$-derivative. So we map
\beq
f  \rightarrow \left[\frac{(1+\mu^z)}{2} + \frac{(1-\mu^z)}{2} \tau^y\right] f \,.
\eeq
Then the Lagrangian density becomes
\beq
\mathcal{L} = f^\dagger \left[  \partial_\tau  - 2i t_1 \tau^x \partial_x - 2 i t_1 \tau^y \partial_y  - 4 t_2  \tau^z \right] f
\eeq
We also want to make the matrices associated with derivatives symmetric. So we map
\beq
f \rightarrow \frac{(1 + i \tau^x)}{\sqrt{2}} f
\eeq
to obtain
\beq
\mathcal{L} = f^\dagger \left[  \partial_\tau  - 2i t_1 \tau^x \partial_x - 2 i t_1 \tau^z \partial_y  + 4 t_2  \tau^y \right] f \,.
\eeq
This form establishes the common chirality of both Dirac nodes.

The field-theoretic formulation of the gauge fluctuations about this spinon Hamiltonian have been
discussed extensively in the literature, and a recent discussion is in Ref.~\onlinecite{Wang18}.
The $\pi$-flux state, at $t_2 = 0$, is described by a SU(2)$_{cg}$ gauge theory coupled to 2 species of massless 2-component
Dirac fermions. The Dirac mass term, can be viewed as the condensate of a fluctuating scalar field, $\phi$. 
In this manner, we obtain a continuum relativistic Lagrangian, which we can write schematically as
\beq
\mathcal{L}_\phi = i \overline{f} \gamma_\mu D^a_\mu f - i \lambda \phi \overline{f}f + (\partial_\mu \phi)^2 + s \phi^2 + u \phi^4 \,.
\label{Lphi}
\eeq
Here $\gamma_\mu$ are the Dirac gamma matrices, $D^a_\mu$ is a co-variant derivative of a SU(2)$_{cg}$ gauge field $a$, 
and $\lambda$ is a Yukawa coupling
to the real scalar field $\phi$. When $\langle \phi \rangle = 0$, we obtain the $\pi$-flux gapless spin liquid: Ref.~\onlinecite{Wang18} argued
that this spin liquid describes the phase transition between the N\'eel ordered and VBS phases. 
As we tune the scalar mass $s$, we undergo a quantum phase transition to a phase which spontaneously breaks time-reversal symmetry with $\langle \phi \rangle \neq 0$: this is the chiral spin liquid. The $\phi$ condensate gives the fermions a mass, and integrating out the massive fermions yields a Chern-Simons term in the SU(2) gauge field \cite{Wang18}.

The field theory in Eq.~(\ref{Lphi}) can also describe a direct phase transition from the chiral spin liquid to a N\'eel ordered phase across a deconfined critical point at $s=s_c$. When $s<s_c$, $\phi$ is condensed, leading to a chiral spin liquid, as noted above. Exactly at $s=s_c$, we have critical $\phi$ fluctuations along with gapless fermions, and we presume this stabilizes a deconfined conformal field theory. For $s>s_c$, $\phi$ is gapped and can be ignored; the remaining theory is SU(2) quantum chromodynamics with $N_f=2$ flavors of two-component massless fermions, and recent Monte-Carlo simulations \cite{Karthik18,Snir18} indicate that such a theory undergoes confinement and chiral symmetry breaking.  A reasonable conclusion is that the chiral symmetry breaking leads to N\'eel order \cite{Snir18}.

We also computed the orbital currents in the one-band chiral spin liquid state above by introducing the charged boson excitations as described below in Section~\ref{sec:piflux3}.
 We confirmed that all currents vanished in this one-band formulation, as stated in Table~\ref{LoopCurrentPattern}. As already mentioned above, this vanishing is related to the four-fold rotational symmetry of pattern D. Given any link oriented from site $i$ to $j$, we can perform a $\pi$-rotation about site $i$, and follow it by a translation, to deduce that the current should be the same on the link oriented from site $j$ to $i$; hence all currents vanish in the one-band model. This argument does not extend to the three-band model because then the O sites are not centers of four-fold rotation symmetry. The following section explicitly demonstrates the presence of orbital currents in the three-band formulation of the chiral spin liquid.

\subsection{Extension to the three-band model}
\label{sec:piflux3}

We will now apply the parameterization in Eq.~(\ref{cbf}) to the three-band model, and deduce the structure of the mean-field theory after factorizing expressions like those in Eq.~(\ref{fbhopping}) into fermion and boson bilinears. All factorizations will conserve spin,
and the boson and fermion numbers separately, as is needed for a theory of an insulator or metal. 
We will also consider here the effective action for the bosons $b_1$ and $b_2$, and use it to compute the orbital charge currents on each link. 

Before describing the structure of the effective Hamiltonians of the fermions and bosons, we first consider the fate of the kinematic Berry phase terms in the Lagrangian. From the definitions in Section~\ref{sec:su2}, we have
\bea\begin{split}
c^\dagger_{i \alpha} \partial_\tau c^\pdagger_{i \alpha} &= \frac{1}{2} \mbox{Tr} \left[ C_i^\dagger \partial_\tau C_i \right]  \\
&= 
\frac{1}{2} \mbox{Tr} \left[ F_i^\dagger \partial_\tau F_i \right] + \frac{1}{2} \mbox{Tr} \left[ R_{ci}^\dagger F_i^\dagger F_i \partial_\tau R_{ci}  \right]  \,, \quad \label{kinfb} \end{split}
\eea
where we have freely integrated by parts. The first term on the right hand side confirms that
the $F_i$ fermions are canonical, as we have already assumed. In the second term, we replace 
the fermion bilinear with an expectation value (in the Grassman path integral), with
$\left \langle F_i^\dagger F_i \right\rangle \propto \sigma^z$. Then using
\beq
\frac{1}{2} \mbox{Tr} \left[ R_{ci}^\dagger \sigma^z \partial_\tau R_{ci}  \right] = b_{i1}^\ast
\partial_\tau b_{i1} + b_{i2}^\ast
\partial_\tau b_{i2} \,, \label{bkin}
\eeq
we confirm that the $b_{i1}$ and $b_{i2}$ behave like canonical bosons (after rescaling). Note that the right-hand-side of Eq.~(\ref{bkin}) is invariant under global SU(2)$_c$, as it must be.

\begin{figure}
\centering
\includegraphics[width=3in]{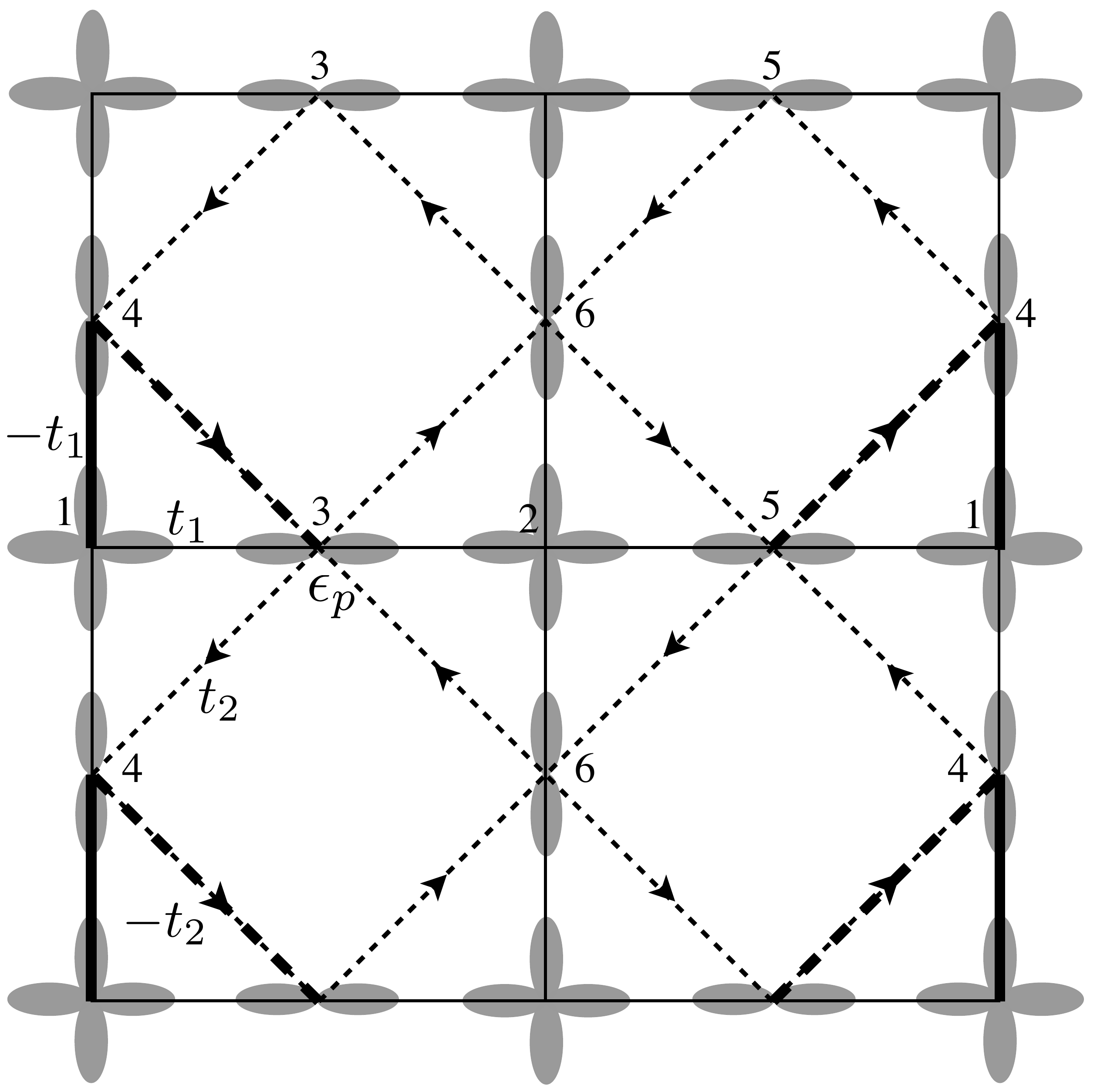}
\caption{Hopping matrix elements of the fermionic spinons, $f$, in the three-band CuO$_2$ model. The on-site energy on the O sites is $\epsilon_p$. 
Thin lines are $t_1$, thick lines are $-t_1$, dashed lines are $t_2$ and thick dashed lines are $-t_2$. The arrows determine the choice between $t_2$ and $t_2^\ast$.
The parameters $t_1$
and $\epsilon_p$ are real, while $t_2$ is complex in the chiral spin liquid. Real $t_2$ yields the $\pi$-flux state.}
\label{fig:threeband}
\end{figure}

Let us now describe the structure of the fermion Hamiltonian. As in Section~\ref{sec:piflux}, this is obtained by factorizing the boson bilinears in Eq.~(\ref{fbhopping}). Then the analog of 
Eq.~(\ref{HfM}) is now
\bea
H_f &=&
\sum_k \sum_{a=3}^6 \epsilon_p f_{a \alpha}^\dagger (k)  f_{a \alpha} (k) \nonumber \\
&~& +  \sum_k \sum_{a,b=1}^6 f_{a \alpha}^\dagger (k) M_{ab} (k) f_{b \alpha} (k) \,, \label{HfM3}
\eea
where $a, b$ extends over the 6 sites in a unit cell illustrated in Fig.~\ref{fig:threeband}.
The matrix $M_{ab}$ contains the hopping matrix elements $t_{1,2}$ as illustrated. The signs of the real $t_1$ are chosen so that there is $\pi$ flux in each square lattice plaquette of Cu atoms. The complex $t_2$ are chosen so that the flux in all loops is invariant under all square lattice translational and rotational symmetries. We numerically diagonalized in Eq.~(\ref{HfM3}) and found that the spectrum was very similar to that of the one-band model in Section~\ref{sec:piflux}. At $t_2=0$, there are 2 massless Dirac nodes. Turning on a non-zero $t_2$ opens up a gap at both nodes with the same chirality, so that the bands near the Dirac nodes have a non-zero Chern number. Consequently, the low-energy theory of the spinons is still given by the field theory in Eq.~(\ref{Lphi}).

Next, we determined the effective Hamiltonian for the bosons by factorizing the fermion bilinears in Eq.~(\ref{fbhopping}).
This yields a Hamiltonian for $b_1$ of the form
\bea
H_{b1} &=&
\sum_k \sum_{a=3}^6 \epsilon_{p1} b_{1a}^\ast (k)  b_{1a} (k) \nonumber \\
&~& +  \sum_k \sum_{a,b=1}^6 b_{1a}^\ast (k) \widetilde{M}_{ab} (k) b_{1b} (k) \,, \label{Hb1}
\eea
where $\widetilde{M}$ is a matrix with same structure as $M$, but with the $t_1$ and $t_2$ replaced by $\widetilde{t}_1$ and $\widetilde{t}_2$. Similarly, the Hamiltonian for $b_2$ is
\bea
H_{b2} &=&
\sum_k \sum_{a=3}^6 \epsilon_{p2} b_{2a}^\ast (k)  b_{2a} (k) \nonumber \\
&~& -  \sum_k \sum_{a,b=1}^6 b_{2a}^\ast (k) \widetilde{M}_{ab} (k) b_{2b} (k) \,, \label{Hb2}
\eea
The condensation of $b_{1,2}$ leads to superconducting states, and so we only consider $H_{b1,2}$ at temperatures above the 
condensation temperature of the bosons. 

Finally, we used the above quadratic Hamiltonian for the fermions and bosons to compute the gauge-invariant charge current
on each link. From Eq.~(\ref{fbhopping}), the expression for the current, $J_{ij}$, on the link connecting sites $i$ and $j$ is 
\beq
J_{ij} = 2 t_{ij} \mbox{Im} \left[
 \left\langle f_{i \alpha}^\dagger f_{j \alpha} \right\rangle \left\langle b_{i1}^\ast b_{j1} \right\rangle -
\left\langle f_{j \alpha}^\dagger f_{i \alpha} \right\rangle \left\langle b_{j2}^\ast b_{i2} \right\rangle \right] \,.
\label{current}
\eeq
We assumed sample values of the parameters $t_{1,2}$, $\widetilde{t}_{1,2}$, $\epsilon_p$, $\epsilon_{p1,2}$, and the fermion and boson chemical potentials. We then verified numerically that for $t_2, \widetilde{t}_2$ complex, the currents $J_{ij}$ 
display pattern D of Table~\ref{LoopCurrentPattern}.

We note that the above formalism can be applied equally to the undoped and doped antiferromagnets, as long as the temperature is high enough so that the bosons are not condensed. As $T \rightarrow 0$, the Chern-Simons term will convert the bosons into semions, and we expect anyonic superconductivity at non-zero doping \cite{RBL88,FHL89,CWWH89}. This should be contrasted with the low temperature metallic states with the same time-reversal and mirror symmetry breaking, but distinct $\mathbb{Z}_2$ topological order, obtained in the formalism of Section~\ref{BosonicSpinons} for pattern D.

%=================================================================================================================== 
\section{Symmetry signatures of different loop currents in magnetic field}\label{SymmetrySignatures}
Having provided various spin-liquid descriptions of loop-current order, we next comment on the distinct symmetry-signatures of the different loop current patterns in \tableref{LoopCurrentPattern}. Based solely on symmetries, the following discussion does not depend upon which spin-liquid description is used. For concreteness, we use the three-orbital model with bosonic spinons, see \secref{ThreeOrbiModLatt}, to illustrate our general symmetry arguments by explicit calculations.

Although the magnetic point group uniquely determines the loop current pattern in \tableref{LoopCurrentPattern}, some experimental probes are only sensitive to time-reversal-invariant observables, such as STM. This is why STM can only be used to extract the (local) magnetic point group modulo time-reversal. For instance, in the case of pattern C and D, all time-reversal-invariant observables will be invariant under all symmetries of the square lattice, \textit{i.e.}, the presence of these types of loop currents is entirely ``invisible'' to STM. For both pattern A and B, only nematic symmetry breaking (reduction of point group from $C_{4v}\rightarrow C_{2v}$) is accessible. Using the overlap between neighboring atoms as examples, the symmetry signatures for time-reversal invariant observables of all patterns are illustrated in \tableref{LoopCurrentPattern}.

However, distinct symmetry signatures of pattern A--D can be revealed in time-reversal symmetric observables once a magnetic field along the $z$ direction is applied, $B_z\neq 0$.
Upon noting that the magnetic field is invariant under $C_4$ and $\Theta\sigma_{xz}$, the resulting symmetry signatures for the local overlaps of the atomic wavefunctions are straightforwardly obtained and summarized in \tableref{LoopCurrentPattern}. 
For instance, an imbalance in the overlap of the Cu-O bonds along the positive and negative $x$- and $y$-directions cannot be directly induced by the magnetic field alone but will be proportional to the product of the magnetic field and the order parameter of loop-current pattern B; the experimental detection of this imbalance in a magnetic field would be strong evidence for loop-currents with the symmetries of pattern B.

We note that pattern D is special as it is the only configuration that transforms exactly as the magnetic field along the $z$ direction and, hence, does not lead to any additional symmetry breaking when an orbital magnetic field $B_z$ is applied. In this sense, it can be regarded as an orbital ferromagnet.
This is also the reason why this pattern exhibits an anomalous Hall and a non-zero Kerr effect as discussed in the introduction. 

To illustrate these symmetry arguments, we consider the kinetic energies ($\vec{A}$ denotes the magnetic vector potential)
\begin{equation}
K_{ij} = - t_{ij} \braket{e^{i \int_{i}^{j}\diff\vec{r}\, \vec{A}(\vec{r})}  \psi^\dagger_{i\alpha} \left(U_{ij}\right)_{\alpha\beta} \psi^\pdagger_{j\beta} + \text{H.c.}}  \label{KinEnergies}
\end{equation}
along both the four different O-O bonds associated with $V_n$ in \figref{PhaseConvetions} (denoted by $K_n^{\text{O-O}}$, $n=1,2,3,4$, in the following) and along the four Cu-O bonds associated with $W_n$ (represented by $K_n^{\text{Cu-O}}$) within the three-orbital model of \secref{ThreeOrbiModLatt}.

In analogy to the derivation of the expectation values (\ref{Currents}) of the currents, obtained by treating the hopping amplitudes $t$, $t'$ as perturbations, we can calculate $K_{ij}$ in the presence of a magnetic field. As shown in \appref{SmallHybr}, the leading terms that depend on the magnetic field read as
\begin{subequations}\begin{align}\begin{split}
& K_n^{\text{O-O}}(B_z) - K_n^{\text{O-O}}(0) \\ & \quad =   t^2 t' \left[f_b  \sin(\phi) b_n+f_e  (1-\cos(\phi)) e_n\right]\end{split}
\end{align}
for the O-O and
\begin{align}\begin{split}
& K_n^{\text{O-Cu}}(B_z) - K_n^{\text{O-Cu}}(0)   \\ & \qquad =   t^2 t' \Bigl[-f_b \sin(\phi)(b_n+b_{n-1}) \\ & \qquad\qquad\quad +f_e  (1-\cos(\phi))(e_n+e_{n-1})\Bigr]\end{split}
\end{align}\label{BondInMagn}\end{subequations}
for the Cu-O bonds. In \equref{BondInMagn}, the dependence on the magnitude $H=|\vec{H}_j|$ of the Higgs-field (independent of $j$ due to translational symmetry), on $\mu$, and $\Delta$ is described by the prefactors $f_{e,b}=f_{e,b}\left(H,\Delta,\mu\right)$, which are given in \equsref{FormOfFb}{FormOfFe}, and $\phi$ is the magnetic flux per elementary Cu-O-O triangle in units of the flux quantum.

From these expressions and taking into account the symmetries of $b_n$ and $e_n$ of the different loop-current patterns, we reproduce all symmetry signatures in the presence of a magnetic field discussed above and shown in \tableref{LoopCurrentPattern}. For instance, while the kinetic energies of pattern D are affected by a magnetic, we see from \equref{BondInMagn} that all bonds are affected equally in that case since $b_n=b_{n+1}$ and $e_n=e_{n+1}$.

We finally point out that, in general, also the symmetries of the loop current patterns themselves are affected by the magnetic field. The generalization of \equref{Currents} to $B_z\neq 0$ reads as
\begin{align}\begin{split}
 J_n^{\text{O-O}}(B_z) &=   t^2 t' \left[f_b  \cos(\phi) b_n - f_e  \sin(\phi) e_n\right], \\
 J_n^{\text{O-Cu}}(B_z) &=   t^2 t' \bigl[f_b  \cos(\phi) (b_{n-1}-b_{n}) \\ & \qquad\quad  + f_e  \sin(\phi) (e_{n}-e_{n-1})\bigr]. 
\end{split}\label{CurrentInMagnField}\end{align}
Note that, despite the reduced symmetries of the current patterns, the currents are still intra-unit-cell loop currents in the sense that \equref{BlochTheorem} is satisfied. 
While the pre-factor $\cos(\phi)$ of $b_n$ in \equref{CurrentInMagnField} leads to a change in magnitude of the loop currents with magnetic flux $\phi$, the terms proportional to $\sin(\phi)$ and $e_n$ describe the change in symmetry of the loop-current pattern. For example, the magnitude of the loop currents of the Cu-O bonds along the positive and negative $x$- and $y$-axis become different for pattern B when a magnetic field is applied; the symmetry changes for all loop current patterns are summarized in \tableref{LoopCurrentPattern}. We finally note that pattern B is the only pattern where the application of a magnetic field $B_z$ leads to a current along a bond [the O-O bonds parallel to the diagonals $x=y$, $J_{2,4}^{\text{O-O}}(B_z)\neq 0$] that has to be zero for $B_z=0$ [$J_{2,4}^{\text{O-O}}(0)= 0$].

%=================================================================================================================== 
\section{Conclusions}
\label{Conclusion}

This paper has described the low-energy theories of various square-lattice spin liquid states which break some combinations of time-reversal and mirror reflection symmetries. We focused on the 4 distinct patterns shown in Table~\ref{LoopCurrentPattern}, and constructed them with the two methods described in Section~\ref{sec:su2}: the $\text{SU}(2)_{sg}$ gauge theory with bosonic spinons in Section~\ref{BosonicSpinons}, and the $\text{SU}(2)_{cg}$ gauge theory with fermionic spinons in Section~\ref{FermionicSpinons}. 
As two out of the four patterns in \tableref{LoopCurrentPattern} only allow for finite orbital currents in the three-orbital model of the CuO$_2$ planes, we have presented realizations in both the one-band and three-band models. 
In addition to the orbital currents themselves, we have also constructed further order parameters for all different patterns in terms of the degrees of freedom of the low-energy theories -- both in the one- and in the three-band model. The scalar spin-chirality operators in \equref{TripleProdcts} assume non-zero expectation values along the elementary Cu-O-O triangles in \figref{PhaseConvetions} with the relative sign determined by the magnetic point group of the respective pattern (see \secref{SpinModels}). The simplest spin-chirality operators in the one-orbital model involve neighboring Cu atoms and are given in \equref{TripleProdcts2}.
Finally, in \secref{SymmetrySignatures}, we have discussed the behavior of the different patterns in the presence of a magnetic field.
We have shown that the magnetic field in conjunction with the non-trivial magnetic point groups of the current patterns lead to unique deformations of the orbital overlap along the elementary Cu-O bonds, which we propose as a possible route towards distinguishing different loop-current patterns experimentally.

\begin{figure}[tb]
\begin{center}
\includegraphics[width=\linewidth]{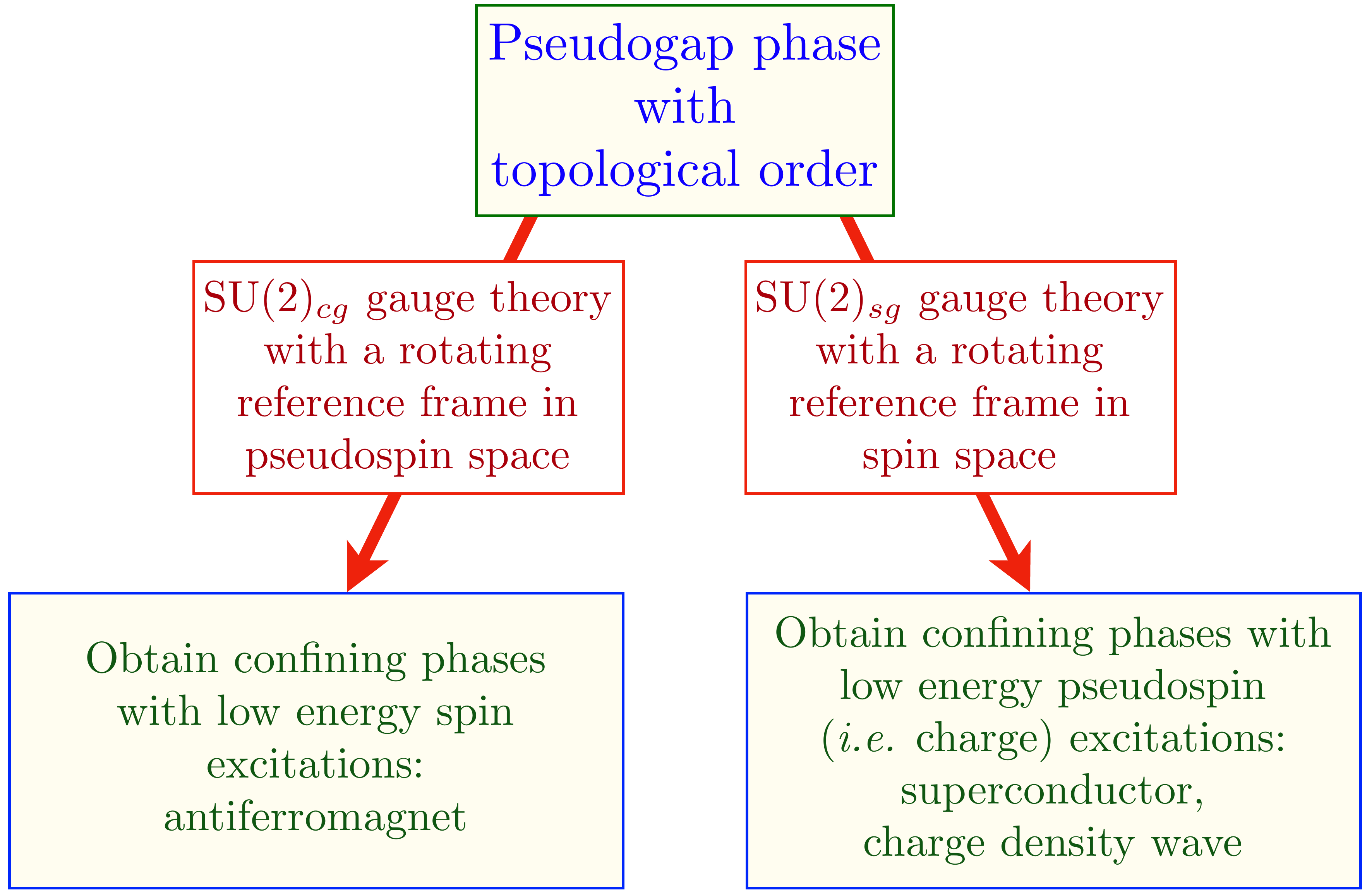}
\caption{Sketch of a proposed relationship of the phases of the cuprates to a ``parent'' pseudogap phase with topological order. This parent phase is presumed to reach confining states via deconfined critical points described by a SU(2) gauge theory. The $\text{SU}(2)_{cg}$ and $\text{SU}(2)_{sg}$ theories reach the phases found
at lower and higher doping, respectively. Note that the confining phase has low energy spin (pseudospin) excitations when we transform to a rotating reference frame in pseudospin (spin) space.} 
\label{Overview}
\end{center}
\end{figure}

After obtaining these spin liquid states and discussing their properties, we can now ask about their possible relevance to the physics of the cuprates. A proposal for such applications is illustrated in \figref{Overview}, drawing upon insights from recent quantum Monte Carlo studies on some other, simpler, spin liquid states \cite{Snir16,Snir18}. 
We view the spin-liquid state as a ``parent'' pseudogap phase with topological order. Such a parent state can be described by Higgsing either the $\text{SU}(2)_{sg}$ or $\text{SU}(2)_{cg}$ gauge theories, and different choices for the Higgs fields lead to the symmetry breaking patterns in Table~\ref{LoopCurrentPattern}. 
Now imagine that the gauge theory undergoes a Higgs-confinement transition across a deconfined critical point where the $\text{SU}(2)$ gauge fields are deconfined. Then, as was found in Refs.~\cite{Snir16,Snir18}, and is illustrated in Fig.~\ref{Overview}, the confining phase has low energy excitations (and possible broken symmetries) in the global symmetry which was not gauged. So a $\text{SU}(2)_{sg}$ gauge theory obtained by transforming to a rotating reference frame in spin space, yields a confining phase with low energy pseudospin excitations. Conversely, a $\text{SU}(2)_{cg}$ gauge theory obtained by transforming to a rotating reference frame in pseudospin space, yields a confining phase with low energy spin excitations. From Fig.~\ref{Overview}, it is clear that moving from the pseudogap to the antiferromagnet at lower doing requires the $\text{SU}(2)_{cg}$ gauge theory in this scenario. On other hand, moving from the pseudogap to the superconductor or charge (or pair) density wave at larger doping requires the $\text{SU}(2)_{sg}$ gauge theory. In both cases, a reasonable scenario is that the time-reversal and mirror plane symmetry breaking patterns of the parent spin-liquid survive vestegially into the confining phase.

In our earlier work with others \cite{sachdev2017insulators,PhysRevLett.119.227002,ATSS18}, we presented a physical motivation 
for the symmetry-breaking patterns A and B in Table~\ref{LoopCurrentPattern}. This motivation arose from a $\text{SU}(2)_{sg}$ theory of the pseudogap, which can describe 
ordering transitions to antiferromagnetically ordered states by the condensation of bosonic spinons (such ordering transitions are distinct from the confinement transitions across deconfined critical points discussed above, and in Fig.~\ref{Overview}). We therefore examined 
time-reversal symmetry breaking $\mathbb{Z}_2$ spin liquid states proximate to the N\'eel ordered state, and found patterns A and B as the most likely candidates. 

The approach described in Fig.~\ref{Overview} indicates an alternate route to selecting the symmetry breaking pattern, and this focuses
attention on pattern D. In our analysis in this paper, we established that pattern D has the same pattern of time-reversal and mirror-plane 
symmetry breaking as the chiral spin liquid \cite{KL87,WWZ89}; and this pattern displays spontaneous orbital currents in the three-band model, but not in the 1-band model. By Fig.~\ref{Overview}, we move towards antiferromagnetically ordered states by a $\text{SU}(2)_{cg}$ theory of the spin liquid.
We begin with the $\pi$-flux $\text{SU}(2)_{cg}$ gauge theory of the insulator, which was described thoroughly in the recent work of Wang {\it et al.} \cite{Wang18}, and reviewed in Section~\ref{sec:piflux}.
Wang {\it et al.} argued that such a theory describes the phase transition between a N\'eel state and a VBS. They also noted the possibility that this theory could lead to a stable chiral spin liquid state, as reviewed above in Eq.~(\ref{Lphi}). Such a state is energetically favorable because the condensation of the field $\phi$ opens a gap in the spinon spectrum. A transition from this chiral spin liquid to a confining state would occur at a point where the mass term $s$, in Eq.~(\ref{Lphi}), is tuned to criticality, and this can stabilize a deconfined critical point, as we discussed below Eq.~(\ref{Lphi}). The confining phase on the other side of this critical point is expected to have N\'eel order.
However, the symmetry breaking pattern D of the chiral spin liquid could naturally persist across this confinement transition to N\'eel order.
In the context of the continuum field theory description in Eq.~(\ref{Lphi}), we need formally irrelevant operators to preserve pattern D
when the $\phi$ condensate disappears.
So we have described a route 
to the confining N\'eel state
co-existing with pattern D. This offers a rationale to obtaining pattern D as a vestigial remnant of a chiral spin liquid character of the pseudogap.

%==================================================================================================================
\section*{Acknowledgements}
We acknowledge many insightful discussions with David Hsieh, Alberto de la Torre Duran, Sergio Di Matteo, and Michael Norman on the possibilities for time-reversal symmetry breaking patterns in antiferromagnets.
We thank Shubhayu Chatterjee, Steve Kivelson, Max Metlitksi, Rhine Samajdar, and Ashvin Vishwanath  for valuable discussions. 
This research was supported by the National Science Foundation under Grant No. DMR-1664842. 
Research at Perimeter Institute is supported by the Government of Canada through Industry Canada and by the Province of Ontario through the Ministry of Research and Innovation. SS also acknowledges support from Cenovus Energy at Perimeter Institute. 
MS acknowledges support from the German National Academy of Sciences Leopoldina through grant LPDS 2016-12.

\newpage

%==================================================================================================================
\appendix
%==================================================================================================================
\section{Current and kinetic energy in the limit of small hybridization}
\label{SmallHybr}
In this appendix, we derive expressions for the currents and kinetic energies in the three-orbital model by treating the hopping matrix elements, $t$, $t'$, as perturbations to the onsite energy scales $\vec{H}_j$ and $\Delta_j$. This is similar to the expansion in large Higgs fields of \refcite{sachdev2017insulators} and related to a $t/U$ expansion in the Hubbard model. One can also illustrate the expansion geometrically: when the current and kinetic energy on a bond $i$-$j$ is expressed in terms of the Higgs-fields $\vec{H}_j$ and the gauge connections $U_{ij}$, invariance under $\text{SU}(2)_{sg}$ demands that every single contribution be of the form 
\begin{equation}
  \, \text{Tr} \left( U_{i k_1} (\vec{H}_{k_1}\cdot\vec{\sigma})^{p_1}U_{k_1k_2} \cdots  (\vec{H}_{k_n}\cdot\vec{\sigma})^{p_n}U_{k_nj} \right), \label{GenWilsonLoop}
\end{equation}
%\begin{equation}
%\sum_{a_1, \dots,  a_n=1}^3 H^{a_1}_{k_1}\cdots H^{a_n}_{k_n} \, \text{tr} \left( U_{i k_1} (\sigma_{a_1})^{p_1}U_{k_1k_2} \cdots  (\sigma_{a_n})^{p_n}U_{k_nj} \right),
%\end{equation}
where $p_n = 0,1$ and $(\vec{H}\cdot\vec{\sigma})^0 = \sigma_0$ is understood.
The expression (\ref{GenWilsonLoop}) can be seen as a Wilson loop of length $l=n+2$ with $p_1+p_2 + \dots p_n$ additional Higgs-field insertions and the expansion presented in this appendix is an expansion in the length $l$ of the loops. We will see that the first non-trivial contributions to the currents and kinetic energies show up at order $l=3$ and the associated Wilson-loop-like operators are $b_n$ and $e_n$ as defined in \equref{LeadingWilsonLoops} of the main text. 

To make the presentation compact, we will directly analyze the general case of finite magnetic field where the chargon Hamiltonian is given by \equref{EffectiveChargonHam} with minimal substitution,
\begin{equation}
t_{ij}\, \rightarrow \, t_{ij} \exp\left(i \int_{i}^{j}\diff\vec{r}\, \vec{A}(\vec{r}) \right), \label{MinimalSubst}
\end{equation}
where $\vec{A}(\vec{r})$ denotes the magnetic vector potential. 

At zeroth order in the perturbative expansion, the Green's function $G_{ij}(\tau) = - \braket{T_\tau (\psi^\pdagger_i(\tau)\psi_j^\dagger(0))}$ of the chargon Hamiltonian, where $T_\tau$ is the ordering operator in imaginary time $\tau$,  is purely local and reads
\begin{equation}
G^0_{ij}(i \omega_n) = \delta_{ij} \frac{i\omega_n - \Delta_j + \mu - \vec{H}_j \cdot \vec{\sigma}}{(i\omega_n - \Delta_j + \mu)^2 - \vec{H}_j^2} \label{ZerothOrderGreens}
\end{equation}
in Matsubara representation [$\omega_n =  (2n+1) \pi T$]. The corrections to the Green's function arising from finite hybridization and magnetic field, \textit{i.e.}, from terms in the chargon Hamiltonian $H_{\psi}$ proportional to $t_{ij} e^{i \int_{i}^j \diff \vec{r} \vec{A}} =: \widetilde{t}_{ij}(B)$, is taken into account order by order in the Dyson equation,
\begin{align}\begin{split}
& G_{ij}(i \omega_n) = G^0_{ij}(i \omega_n) - G^0_{ii}(i \omega_n)\widetilde{t}_{ij} U_{ij}G^0_{jj}(i \omega_n) \\ &+ \sum_k G^0_{ii}(i \omega_n)\widetilde{t}_{ik} U_{ik} G^0_{kk}(i \omega_n)\widetilde{t}_{kj} U_{kj}G^0_{jj}(i \omega_n) + \dots\, .\label{DysonExpanded}\end{split}
\end{align}
In order to calculate both the kinetic energies $K_{ij}$, \equref{KinEnergies}, as well as the currents $J_{ij}$, \equref{FormOfCurOp}, simultaneously, let us first investigate the complex expectation values
\begin{equation}
T_{ij} = \widetilde{t}_{ij} \braket{\psi_i^\dagger U_{ij}\psi_j}, \quad i\neq j, \label{DefinitionOfTij}
\end{equation}
which are invariant under both emergent and electromagnetic gauge transformations. $T_{ij}$ in \equref{DefinitionOfTij} is related to the quantities of interest via $K_{ij}=-2\,\Re \,T_{ij}$ and $J_{ij}=2\, \Im \,T_{ij}$ and given by
\begin{equation}
T_{ij} = T \sum_{\omega_n} \widetilde{t}_{ij}\,\text{Tr} \left( U_{ij} G_{ji} (i\omega_n)\right) e^{i\omega_n 0^+}
\end{equation}
in terms of the Matsubara Green's function defined above. Inserting the expansion (\ref{DysonExpanded}), we obtain $T_{ij}$ as a power series in $t$, $t'$ the leading non-trivial term of which we will calculate in the following. Clearly, the zeroth order term in \equref{DysonExpanded} does not give rise to a contribution to $T_{ij}$ as $G^0$ is entirely local. Also the second order term is not of interest to us for the following reasons: First,  $\widetilde{t}_{ij}$ only appears in the combination $\widetilde{t}_{ij}\widetilde{t}_{ji}= t^2_{ij}$ such that there is no dependence on the magnetic field in this order and, second,  $\sum_{\omega_n} \text{Tr} \left( U_{ij} G^0_{jj} U_{ji} G^0_{ii}\right)$ is readily seen to be real implying no contribution to the current. The leading relevant contribution turns out to be the second order term in \equref{DysonExpanded} corresponding to loops of length $l=3$. Upon noting that the three-orbital model, introduced in \secref{ThreeOrbiModLatt} of the main text, only allows for loops along the  elementary Cu-O-O triangles (four of which are shown in \figref{PhaseConvetions}), it can be written as $T_{ij}^{(2)}=t^2t' \sum_k \Delta_{ijk}$, where the sum involves the site/sites $k$ that is/are connected to both $i$ and $j$ via hopping $t$ or $t'$ and
\begin{align}\begin{split}
&\Delta_{ijk} = e^{i\phi_{ijk}} \\ &\quad \times  \int\frac{\diff\omega}{2\pi} \text{Tr}\left(G^0_{ii}(i\omega)U_{ij}G^0_{jj}(i\omega)U_{jk}G_{kk}^0(i\omega)U_{ki} \right).
\end{split}\end{align}
Here $\phi_{ijk}$ is the flux (in units of the flux quantum) through the Cu-O-O triangle $i\rightarrow j \rightarrow k \rightarrow i$ and the temperature has been set to zero for simplicity. As $\Delta_{ijk}$ is invariant under cyclic permutation of the indices, we can, without loss of generality, assume that $i$ is on a Cu atom while $j$ and $k$ are O sites and obtain all other combinations by permutation of the indices. Using the explicit form (\ref{ZerothOrderGreens}) of the zeroth order Green's function, one finds after integration over frequency
\begin{equation}
\Delta_{ijk} = \frac{e^{i\phi_{ijk}}}{2}\left( f_e\left(|\vec{H}_i|,\Delta,\mu\right) e_{ijk} - i f_b\left(|\vec{H}_i|,\Delta,\mu\right) b_{ijk} \right), \label{DeltaExplicit}
\end{equation}
where 
\begin{align}
e_{ijk} &= \text{Tr}\left(U_{ij}U_{jk}U_{ki}\right), \,\,  b_{ijk} = i \,\text{Tr}\left(\vec{H}_i\cdot\vec{\sigma}U_{ij}U_{jk}U_{ki}\right).
\end{align}
As anticipated, $e_{ijk}$ and $b_{ijk}$ are special cases of \equref{GenWilsonLoop} with zero and one Higgs insertion and trivially related to $e_n$ and $b_n$ introduced in \equref{LeadingWilsonLoops}. The functional form of the prefactors $f_b$ and $f_e$ is given by 
\begin{align}
 f_b(H,\Delta,\mu) = \begin{cases} \frac{1}{H(H+s_{\Delta\mu}\Delta)^2}, & H > |\mu|,  \\ \frac{-4 \Delta s_{\Delta\mu}}{\left(\Delta^2-H^2\right)^2}\theta(-s_{\Delta\mu}\mu), & H < |\mu|,   \end{cases}\label{FormOfFb}
\end{align}
and
\begin{align}
 f_e(H,\Delta,\mu) = \begin{cases} \frac{-s_{\Delta\mu}}{(H+s_{\Delta\mu}\Delta)^2}, & H > |\mu|,  \\ \frac{-2  s_{\Delta\mu}\left(H^2+\Delta^2\right)}{\left(H^2-\Delta^2\right)^2}\theta(-s_{\Delta\mu}\mu), & H < |\mu|,   \end{cases}\label{FormOfFe}
\end{align}
respectively, where $s_{\Delta\mu} = \sign(\mu+\Delta)$ and $\theta$ denotes the Heaviside step function. 
From \equref{DeltaExplicit} we immediately find $T^{(2)}$ and, by taking its real and imaginary parts, the expressions (\ref{Currents}), (\ref{BondInMagn}), and (\ref{CurrentInMagnField}) of the main text.

%==================================================================================================================
\section{Possible ans\"atze with $U_{ij} = \mathds{1}$}
\label{PossibleHiggsPhases}
In this appendix, we will prove that neither the three-orbital, \equref{EffectiveChargonHam}, nor the one-orbital chargon model, \equref{EffectiveChargonHamOneOrbital}, allows for an ansatz with $U_{ij} = \mathds{1}$ on all bonds and the symmetries of pattern C and D.  

Due to $U_{ij} = \mathds{1}$, all gauge transformations $G_g(j)$ accompanying a physical symmetry transformation $g$ must be global, $G_g(j)=G_g$. Denoting the (adjoint) representation of $G_g$ on the Higgs field $\vec{H}_j$ by $\mathcal{R}_g \in \text{SO}(3)$, invariance under translation, $g=T_\mu$, along $\mu=x$ and $\mu=y$, allows to write
\begin{equation}
    \vec{H}_{j} = \left(\mathcal{R}_{T_x}\right)^{j_x}\left(\mathcal{R}_{T_y}\right)^{j_y} \vec{H}_{(0,0)}. \label{GeneralFormOfHiggs}
\end{equation}
By virtue of forming a representation, $\mathcal{R}_{T_x}$ and $\mathcal{R}_{T_y}$ must commute. Consequently, there are only two options: \textit{(i)} $\mathcal{R}_{T_x}$ and $\mathcal{R}_{T_y}$ are $\text{SO}(3)$ rotations (with arbitrary angles, denoted by $Q_x$ and $Q_y$ in the following) along the same direction or \textit{(ii)} $\pi$ rotations along two orthogonal directions.

To begin with case \textit{(i)}, we first note that a global gauge transformation allows to choose the common rotational axis to be along $\vec{e}_z$ and to set $\vec{H}_{(0,0)} = (H_0,0,\epsilon)^T$ without loss of generality. \equref{GeneralFormOfHiggs} then becomes  
\begin{equation}
      \vec{H}_j =  \left(H_0\cos(\vec{Q}\vec{r}_j) ,H_0\sin(\vec{Q}\vec{r}_j),\epsilon \right)^T, \label{Option1ForHi}
\end{equation}
which has the form of a conical spiral. As discussed in \secref{PossibleAnsaetze} of the main text and in \refcite{PhysRevLett.119.227002}, it allows to describe pattern A and B. 

In order to represent the symmetries of C and D, the ansatz has to preserve $C_2$. Due to $\vec{H}_{-j} = \text{diag}(1,-1,1)\vec{H}_{j}$ in \equref{Option1ForHi}, this requires that at least one of the three components of $\vec{H}_j$ be zero for all $j$, i.e., either $\epsilon=0$ (planar spiral), $H_0=0$ (ferromagnetic ansatz), or $\sin(\vec{Q}\vec{r}_j) =0$ ($Q_\mu \in \pi \mathbb{Z}$, corresponding to ferro- or antiferromagnetic ansatz).
However, if $\vec{H}_j$ is co-planar, time-reversal will automatically be preserved as $\vec{H}_j \rightarrow -\vec{H}_j$ can be compensated by a global $\pi$ rotation perpendicular to the plane of $\vec{H}_j$ and the symmetries of pattern C and D cannot be represented.

It is finally left to show that the same holds for option \textit{(ii)}. Without loss of generality, we can assume that $\mathcal{R}_{T_x}$ and $\mathcal{R}_{T_y}$ are $\pi$ rotations along the $\vec{e}_x$ and $\vec{e}_y$ direction, respectively. The Higgs field in \equref{GeneralFormOfHiggs} then assumes the form
 \begin{equation}
     \vec{H}_j = \left(  (-1)^{j_y} H_x,  (-1)^{j_x}H_y,   (-1)^{j_x+j_y} H_z \right)^T. \label{TextureOption2}
 \end{equation}
While this Higgs-field texture will automatically preserve $C_2$ and will break time-reversal symmetry (as long as $H_x,H_y,H_z \neq 0$), it cannot yield the correct symmetries since it will necessarily break both $C_4$ and $\Theta C_4$: Under $C_4$, the Higgs field in \equref{TextureOption2} transforms as $\vec{H}_j \rightarrow \text{diag}((-1)^{j_x+j_y},(-1)^{j_x+j_y},1) \vec{H}_j$ which cannot be compensated by a global gauge transformation. The same holds for $\Theta C_4$. Taken together, it is not possible to realize the symmetries of pattern C and D if $U_{ij}=\mathds{1}$.

%==================================================================================================================
\section{Further quartic Higgs terms}
\label{FurtherHiggsTerms}
In this section of the appendix, we show that there are no U(1) symmetric or charge-$4$ quartic terms with two or fewer derivatives that, combined with the quadratic \CP action $\mathcal{S} + \int \diff^2 x \, \diff t \,\mathcal{L}_{P,Q}$ defined in \equsref{StandardCP1Action}{LH}, can give rise to the symmetries of the current pattern C or D.

We begin with U(1) symmetric terms. Spin-rotation symmetry allows for the two different forms
\begin{subequations}\begin{align}
  (z^*_\alpha \partial_{\mu} z^\pdagger_\alpha) \, (z_\beta^* \partial_{\mu'} z^\pdagger_{\beta}),  \\
  (z^*_\alpha \vec{\sigma}_{\alpha\alpha'} \partial^n z^\pdagger_{\alpha'}) \cdot (z_\beta^* \vec{\sigma}_{\beta\beta'}  \partial^{n'} z^\pdagger_{\beta'}).
\end{align}\label{FurtherFourBosonTerms}\end{subequations}
Clearly, we need an even number of spatial derivatives to preserve two-fold rotation symmetry. Without any spatial derivatives, we cannot describe the broken reflection symmetries of the loop current patterns and, hence, the minimal number of spatial derivatives is two. Also condensing $P$ does not break time-reversal symmetry since the terms in \equref{FurtherFourBosonTerms} are U(1) symmetric. Consequently, more than two derivatives are required to obtain the symmetries of pattern C or D with U(1) symmetric quartic terms.

We next consider charge-$4$ four-boson interactions with at most two derivatives. Again, due to spin-rotation symmetry, there are in principle two types of terms,
\begin{subequations}\begin{align}
(\varepsilon_{\alpha\alpha'} z_\alpha^\pdagger \partial_{\mu} z^\pdagger_{\alpha'})\, (\varepsilon_{\beta\beta'} z_\beta^\pdagger \partial_{\mu'} z^\pdagger_{\beta'}),  \\
(\varepsilon_{\alpha\alpha'}\vec{\sigma}_{\alpha'\alpha''} z_\alpha^\pdagger \partial_{\mu} z^\pdagger_{\alpha''})\cdot (\varepsilon_{\beta\beta'}\vec{\sigma}_{\beta'\beta''}  z_\beta^\pdagger \partial_{\mu'} z^\pdagger_{\beta''}).
\end{align}\label{Z4Terms}\end{subequations}
Here we have already used that
\begin{equation}
(\varepsilon_{\alpha\alpha'}\vec{\sigma}_{\alpha'\alpha''} z_\alpha^\pdagger \partial^n z^\pdagger_{\alpha''})\cdot (\varepsilon_{\beta\beta'}\vec{\sigma}_{\beta'\beta''}  z_\beta^\pdagger  z^\pdagger_{\beta''}) = 0.
\end{equation}
The number of options is further reduced by noting that the two terms in \equref{Z4Terms} are actually equivalent.

For the same reason as in \secref{PatternCAndD} of the main text, we need two spatial derivatives in order to be consistent with rotational symmetry $C_4$ (or $\Theta C_4$) and translational invariance while being capable of breaking reflection symmetries at the same time. We are thus left with the single term
\begin{subequations}\begin{align}
(\varepsilon_{\alpha\alpha'} z_\alpha^\pdagger \partial_{a} z^\pdagger_{\alpha'})\, (\varepsilon_{\beta\beta'} z_\beta^\pdagger \partial_{b} z^\pdagger_{\beta'}).
\end{align}\end{subequations}
Condensing $P$ to break time-reversal symmetry does not leave any combination of $\Theta$ and rotations or reflections as residual symmetries and, therefore, cannot give rise to the magnetic point symmetries of pattern C and D. 
   
The only remaining possibility is to condense another time-reversal-odd charge-$4$ Higgs field. However, there is no finite charge-$4$ term with a single derivative (since $\varepsilon_{\beta\beta'} z_\beta^\pdagger  z^\pdagger_{\beta'} =0$) and, as already discussed above, adding a spatial derivative, \textit{i.e.}, $\mu=t,\mu'=x,y$ in \equref{Z4Terms}, will either break the rotation symmetries $C_4$ and $\Theta C_4$ or translational invariance.

%==================================================================================================================
\section{O(3) non-linear sigma model}
\label{NonLinearSigmaModel}
In this appendix, we derive the representation of the order parameter for pattern D in the semi-classical O(3) non-linear sigma-model description of quantum-fluctuating antiferromagnetism. In this approach, the spin $\hat{\vec{S}}_i$ is expressed in terms of the local N\'eel order parameter, $\vec{n}(\vec{x},t)$, obeying $\vec{n}^2=1$, and the canonically conjugate uniform magnetization density $\vec{L}(\vec{x},t)$, with $\vec{n}\cdot \vec{L} = 0$, according to \cite{ssbook}
\begin{equation}
    \hat{\vec{S}}_i = S (-1)^{i_x+i_y} \vec{n}_i \sqrt{1-\hat{\vec{L}}_i^2/S^2} +\vec{L}_i, \label{SemiClassExp}
\end{equation}
where  $\vec{n}_i = \vec{n}(i_x\vec{e}_x+i_y\vec{e}_y,t)$ and analogously for $\vec{L}_i$. Furthermore, the theory has been formally extended to a system of spin-$S$ particles to allow for a systematic semi-classical expansion.

We insert \equref{SemiClassExp} into $ \widetilde{\mathcal{O}}^{S}_D$ defined in \equref{OneOrbitalOP} and expand to leading non-trivial order in gradients and powers of $\vec{L}$. Due to the alternating term $(-1)^{i_x+i_y}$ in \equref{SemiClassExp}, only terms with odd powers of $\vec{L}$ can contribute in the continuum limit. 
We note that, contrary to naive expectations and very similar to our discussion of the \CP description in \secref{PatternCAndD}, there is no contribution with two gradients and one $\vec{L}$: the only term compatible with reflection and rotation symmetries is $\vec{L} \cdot (\partial_x \vec{n} \times \partial_y \vec{n})$, which, however, vanishes identically ($\vec{L}$ and $\partial_a \vec{n}$ are perpendicular to $\vec{n}$ and, hence, co-planar). We note in passing that this is different for pattern C since $\vec{L}\cdot ( \vec{n} \times (\partial_x^2 - \partial_y^2)\vec{n} )$ is not identically zero and has the same magnetic point symmetries as pattern C.

The leading non-trivial contribution to $ \widetilde{\mathcal{O}}^{S}_D$, thus, starts at higher order and reads as
\begin{align}\begin{split}
    \mathcal{O}_D^S \sim & 2S^2 \int\diff^2x\diff t \\ 
    &\, \vec{L}\, \cdot \Biggl[ \frac{1}{3}\left( (\partial_y^3 \vec{n}) \times (\partial_x  \vec{n}) - (\partial_x^3 \vec{n}) \times (\partial_y \vec{n} ) \right) \\ & \quad\qquad  +  (\partial_y\partial_x^2 \vec{n}) \times (\partial_x \vec{n}) - (\partial_x\partial_y^2 \vec{n}) \times (\partial_y \vec{n} )  \\ &\quad\qquad +   (\partial_x\partial_y \vec{n})\times ((\partial_x^2-\partial_y^2)\vec{n}) \\
   &\quad\qquad + \frac{2}{S^2} (\partial_x \vec{L}) \times (\partial_y \vec{L}) \Biggr]. \label{ExpressionForOP}
\end{split}\end{align}
Recalling that $\vec{L}$ transforms as $\vec{n}\times \partial_t \vec{n}$, these terms have very similar structure to the order parameters of the \CP theory discussed in \secref{PatternCAndD}. To be more explicit, consider the second line of \equref{ExpressionForOP} which becomes 
\begin{equation}
    (\vec{n}\cdot \partial_x^3\vec{n}) \, \left[ (\partial_t \vec{n})\cdot (\partial_y \vec{n}) \right] - \left( x\leftrightarrow y\right) \label{SingleTermConsidered}
\end{equation}
after replacing $\vec{L}$ by $\vec{n}\times \partial_t \vec{n}$. Using $\vec{n}=z^\dagger \vec{\sigma} z$, \equref{SingleTermConsidered} contains terms proportional to the terms involving $X_{xy}^x$ and $X_{yx}^y$ in the expression of $\mathcal{O}^{\CPm}_D$ in \equref{OCP1D} which is readily seen by noting
\begin{align}\begin{split}
    (\partial_t \vec{n})\cdot (\partial_a \vec{n}) &= 2 (\varepsilon_{\alpha\alpha'} z_\alpha \partial_t z_{\alpha'}) (\varepsilon_{\beta\beta'} z^*_\beta \partial_a z^*_{\beta'}) + \text{H.c.}, \\
    (\vec{n}\cdot \partial_a^3\vec{n}) &= 2 z^*_\alpha \partial^3_a z_\alpha + 6 (z^*_\alpha \partial_a^2 z_\alpha) (\partial_a z^*_\alpha)z_\alpha + \text{H.c.}.\end{split}
\end{align}

In summary, we have explicitly connected the order parameter $\widetilde{\mathcal{O}}^{S}_D$ in \equref{OneOrbitalOP}, expressed in terms of physical spin operators, to the order parameter $\mathcal{O}^{\CPm}_D$, expressed in terms of charge-$2$ Higgs fields of the \CP theory. Furthermore, the non-linear sigma model analysis has provided a more transparent reason for why the \CP description of pattern D involves five (instead of three as is the case for pattern C) derivatives.

%==================================================================================================================
%\bibliography{orbital.bib}
%merlin.mbs apsrev4-1.bst 2010-07-25 4.21a (PWD, AO, DPC) hacked
%Control: key (0)
%Control: author (72) initials jnrlst
%Control: editor formatted (1) identically to author
%Control: production of article title (1) required
%Control: page (0) single
%Control: year (1) truncated
%Control: production of eprint (0) enabled
%

\end{document}